\documentclass[twoside,twocolumn,9pt]{article}
\usepackage{url}
\usepackage{extsizes}
\usepackage[super,sort&compress,comma]{natbib} 
\usepackage[version=3]{mhchem}
\usepackage[left=1.5cm, right=1.5cm, top=1.785cm, bottom=2.0cm]{geometry}
\usepackage{balance}
\usepackage{mathptmx}
\usepackage{amssymb}
\usepackage{amsmath}
\usepackage{bm} 
\usepackage{sectsty}
\usepackage{graphicx} 
\usepackage{lastpage}
\usepackage[format=plain,justification=justified,singlelinecheck=false,font={stretch=1.125,small,sf},labelfont=bf,labelsep=space]{caption}
\usepackage{float}
\usepackage{fancyhdr}
\usepackage{fnpos}
\usepackage[english]{babel}
\addto{\captionsenglish}{%
  \renewcommand{\refname}{Notes and references}
}
\usepackage{array}
\usepackage{droidsans}
\usepackage{charter}
\usepackage[T1]{fontenc}
\usepackage[usenames,dvipsnames]{xcolor}
\usepackage{setspace}
\usepackage[compact]{titlesec}
\usepackage{hyperref}
\definecolor{cream}{RGB}{222,217,201}

\begin{document}
\pagestyle{fancy}
\thispagestyle{plain}
\fancypagestyle{plain}{
\renewcommand{\headrulewidth}{0pt}
}

\makeFNbottom
\makeatletter
\renewcommand\LARGE{\@setfontsize\LARGE{15pt}{17}}
\renewcommand\Large{\@setfontsize\Large{12pt}{14}}
\renewcommand\large{\@setfontsize\large{10pt}{12}}
\renewcommand\footnotesize{\@setfontsize\footnotesize{7pt}{10}}
\makeatother

\renewcommand{\thefootnote}{\fnsymbol{footnote}}
\renewcommand\footnoterule{\vspace*{1pt}%
\color{cream}\hrule width 3.5in height 0.4pt \color{black}\vspace*{5pt}} 
\setcounter{secnumdepth}{5}

\makeatletter 
\renewcommand\@biblabel[1]{#1}            
\renewcommand\@makefntext[1]%
{\noindent\makebox[0pt][r]{\@thefnmark\,}#1}
\makeatother 
\renewcommand{\figurename}{\small{Fig.}~}
\sectionfont{\sffamily\Large}
\subsectionfont{\normalsize}
\subsubsectionfont{\bf}
\setstretch{1.125} 
\setlength{\skip\footins}{0.8cm}
\setlength{\footnotesep}{0.25cm}
\setlength{\jot}{10pt}
\titlespacing*{\section}{0pt}{4pt}{4pt}
\titlespacing*{\subsection}{0pt}{15pt}{1pt}

\fancyfoot{}
\fancyfoot[RO]{\footnotesize{\sffamily{1--\pageref{LastPage} ~\textbar  \hspace{2pt}\thepage}}}
\fancyfoot[LE]{\footnotesize{\sffamily{\thepage~\textbar\hspace{3.45cm} 1--\pageref{LastPage}}}}
\fancyhead{}
\renewcommand{\headrulewidth}{0pt} 
\renewcommand{\footrulewidth}{0pt}
\setlength{\arrayrulewidth}{1pt}
\setlength{\columnsep}{6.5mm}
\setlength\bibsep{1pt}

\makeatletter 
\newlength{\figrulesep} 
\setlength{\figrulesep}{0.5\textfloatsep} 

\newcommand{\topfigrule}{\vspace*{-1pt}%
\noindent{\color{cream}\rule[-\figrulesep]{\columnwidth}{1.5pt}} }

\newcommand{\botfigrule}{\vspace*{-2pt}%
\noindent{\color{cream}\rule[\figrulesep]{\columnwidth}{1.5pt}} }

\newcommand{\dblfigrule}{\vspace*{-1pt}%
\noindent{\color{cream}\rule[-\figrulesep]{\textwidth}{1.5pt}} }

\makeatother

\twocolumn[
  \begin{@twocolumnfalse}
\vspace{1em}
\sffamily
\begin{tabular}{m{0.5cm} p{16cm} }

& \noindent\LARGE{\textbf{Non-equilibrium Growth and Twist of Cross-linked Collagen Fibrils}} \\
\vspace{0.3cm} & \vspace{0.3cm} \\
& \noindent\large{Matthew P Leighton, Laurent Kreplak, and Andrew D Rutenberg$^{\ast}$} \\
& \today \\

& \noindent\normalsize{The lysyl oxidase (LOX) enzyme that catalyses cross-link formation during the assembly of collagen fibrils \textit{in vivo} is too large to diffuse within assembled fibrils, and so is incompatible with a fully equilibrium mechanism for fibril formation. We propose that enzymatic cross-links are formed at the fibril surface during the growth of collagen fibrils; as a consequence no significant reorientation of previously cross-linked collagen molecules occurs inside collagen fibrils during fibril growth \textit{in vivo}. By imposing local equilibrium only at the fibril surface, we develop a coarse-grained quantitative model of \textit{in vivo} fibril structure that incorporates a double-twist orientation of collagen molecules and a periodic D-band density modulation along the fibril axis. Radial growth is controlled by the concentration of available collagen molecules outside the fibril. In contrast with earlier equilibrium models of fibril structure, we find that all fibrils can exhibit a core-shell structure that is controlled only by the fibril radius. At small radii a core is developed with a linear double-twist structure as a function of radius. Within the core the double-twist structure is largely independent of the D-band.  Within the shell at larger radii, the structure approaches a constant twist configuration that is strongly coupled with the D-band. We suggest a stable radius control mechanism that corneal fibrils can exploit near the edge of the linear core regime; while larger tendon fibrils  use a cruder version of growth control that does not select a preferred radius.
} 
\end{tabular}
 \end{@twocolumnfalse} \vspace{0.6cm}
 ]
\renewcommand*\rmdefault{bch}\normalfont\upshape
\rmfamily
\section*{}
\vspace{-1cm}

\footnotetext{Department of Physics and Atmospheric Science,
              Dalhousie University, Halifax,
              Nova Scotia, Canada B3H 4R2. }
\footnotetext{\textit{$^{\ast}$~Corresponding author, email: andrew.rutenberg@dal.ca }}

\section{Introduction} 
The abundant and diverse family of collagen molecules \cite{Ricard-Blum:2011} speaks to the diversity of collagenous structures that are found in nature, and their importance in organism function. Mechanically-loaded collagenous tissues such as bone, tendon, cartilage, and skin are hierarchically assembled materials.\cite{Sherman:2015} Collagen fibrils are  important components of these tissues.

Collagen fibrils are long and approximately cylindrical, and are made up of many long, chiral, semi-flexible collagen molecules.\cite{Rezaei:2018} A robust and characteristic periodic density modulation along the length of the fibril is called the D-band.\cite{Fang:2012, Quan:2015} While the collagen molecules are almost parallel to the fibril axis, they exhibit a  tilt with respect to the axis that is visible both at the fibril surface \cite{Raspanti:2018} and in the interior \cite{Holmes:2001, Bell:2018} -- and that is particularly pronounced in corneal fibrils.  Corneal fibrils also exhibit tight radius control -- which is necessary for corneal transparency.\cite{Meek:2015} Conversely, tendon exhibits an exceptionally broad range of fibril radii,\cite{Parry:1978, PattersonKane:1997, Goh:2012, Kalson:2015} and a much lower molecular tilt.\cite{Reale:1981, Raspanti:2018}

Atomic-scale molecular dynamics (MD) simulations using untilted model fibrils have confirmed the equilibrium stability of the D-band.\cite{Zhou:2016, Xu:2018} Coarse-grained models based on cholesteric liquid-crystals \cite{Brown:2014, Cameron:2018} have demonstrated how tilted fibrils can be thermodynamically stable with respect to a bulk cholesteric phase. These coarse-grained models have been recently extended with phase-field crystal theory to also capture the equilibrium structure of D-banded and titled fibrils.\cite{Cameron:2020} These equilibrium approaches are directly applicable to fibrils that self-assemble \emph{in vitro}.\cite{Harris:2013, Gobeaux:2008} While we expect that fibril assembly  respects these equilibrium energetics, there are non-equilibrium aspects of fibril formation to consider \emph{in vivo} that arise from molecular cross-linking.

Enzymatic cross-links, catalyzed by the enzyme lysyl oxidase (LOX), \cite{Kagan:1991} contribute significantly to tissue elasticity \cite{Eekhoff:2018} -- as demonstrated  experimentally \cite{Makris:2014} and in MD studies.\cite{Depalle:2015}  These cross-links represent covalent bonds, which are not remodelled after they are formed.

The diffusion of LOX into fully grown fibrils appear to be restricted by size constraints. $10$kDa Dextran molecules are excluded from diffusing into hydrated collagen fibrils.\cite{EkaniNkodo:2003} A similar molecular-weight cut-off of various molecules was established by gel-filtration chromatography using a column packed with tendon fibers.\cite{Toroian:2007} LOX, which has a molecular weight of more than $30$kDa,\cite{Kagan:1991} is thus too large to penetrate into assembled fibrils. Supporting this, the radius of gyration of LOX ($3.7$nm)\cite{Vallet:2018} is substantially larger than the narrowest part of subchannels within the fibril structure ($2nm$)\cite{Xu:2018, Xu:2020}. We conclude that enzymatic cross-linking is restricted to the fibril surface; so that cross-links within the fibril must have been made during fibril formation. Supporting this conclusion, enzymatic cross-links  form throughout  fibril growth,\cite{Marturano:2014} and their formation is crucial for proper fibril development \emph{in vivo}.\cite{Herchenhan:2015}

The irreversible formation of cross-links during fibrillogenesis is incompatible with fully equilibrium models of fibril formation. These models allow collagen molecules to freely rearrange throughout the fibril during fibril growth to continually minimize the total free energy of the entire fibril.\cite{Brown:2014, Cameron:2018, Cameron:2020} The incompatibility of cross-linking and equilibrium structure raises fundamental questions for \textit{in vivo} fibrillogenesis: how is fibril radius determined, how does molecular tilt emerge, and how do tilt and D-band structure co-exist within cross-linked fibrils? We address these questions in this paper. 

Radial fibril growth is a slow process \emph{in vivo}: only approximately one molecular layer is added every day during development.\cite{Kalson:2015, Hulmes:1995} As a result, local equilibrium can be achieved at the fibril surface before cross-linking occurs and additional molecular layers bury the cross-linked collagen. With this as motivation, we have developed a new non-equilibrium model for \textit{in vivo} collagen fibril formation. In our model the formation of cross-links during fibrillogenesis freezes the relaxation of internal fibril structure -- but local equilibrium energetics determines structure formation at the fibril surface during growth.

\section{Model}
\begin{figure}[t!] 
\centering
  \includegraphics[height=6cm]{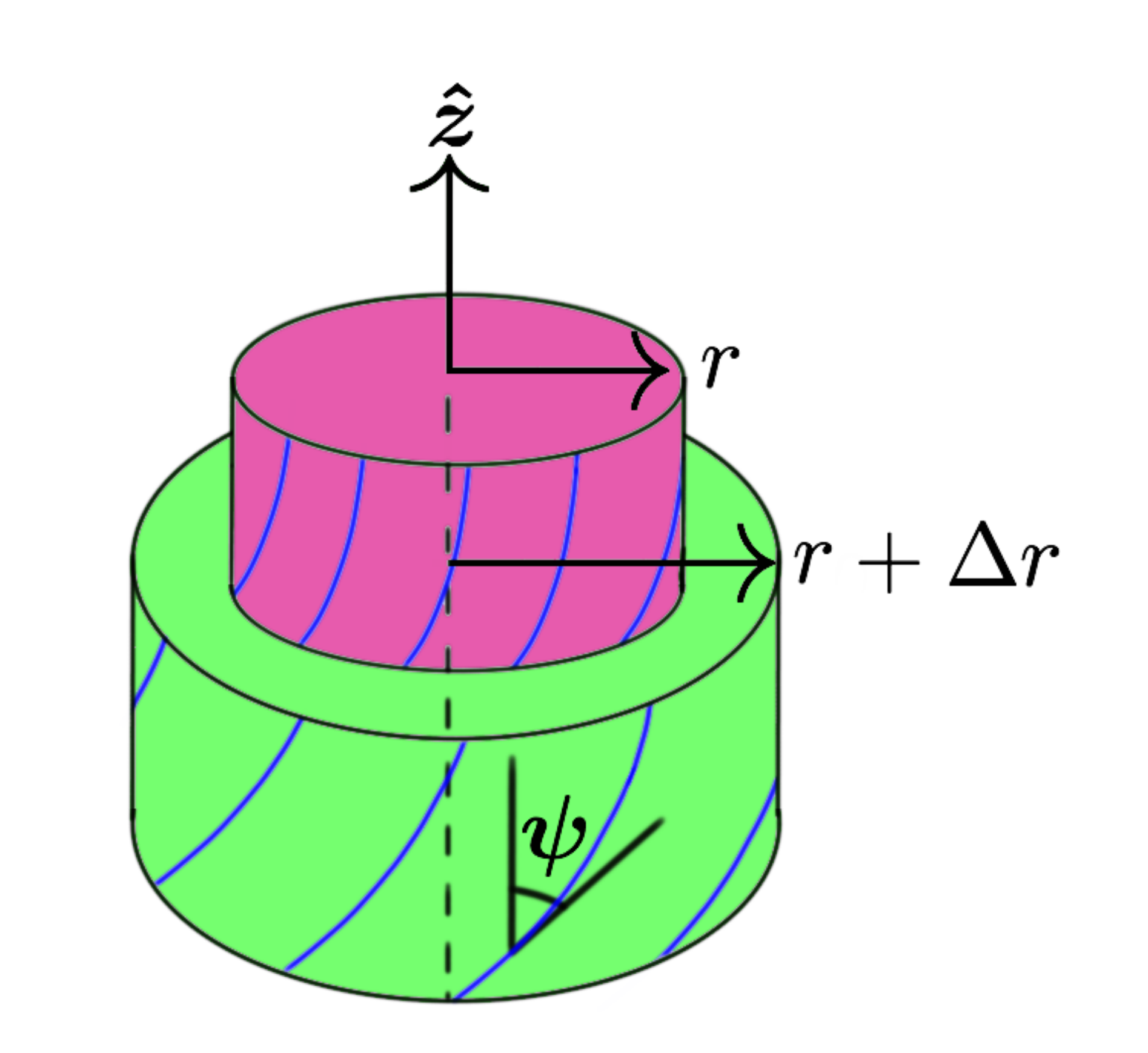}
  \caption{Schematic cutaway of our non-equilibrium fibril model. The surface at $r+ \Delta r$ (in green) is cross-linked while in local equilibrium, while the buried structure below the surface (magenta, at radius $ \leq r$) has frozen structure due to previously formed covalent cross-links. The director field of tilted collagen molecules is represented by curved blue lines, and is characterized by a twist angle $\psi(r)$ with respect to the fibril axis $\hat{z}$. We take the continuum limit $\Delta r \rightarrow 0$.}
  \label{fig:cartoon}
\end{figure}

Our qualitative model is that during radial fibril growth enzymatic cross-linking  freezes the average molecular orientation of interior collagen molecules in place (interior magenta region in Fig.~\ref{fig:cartoon}). Nevertheless, because both the fibril growth and the cross-linking dynamics remain slow compared to timescales of molecular motion, the molecular orientation of surface collagen stays in local equilibrium (surface green region in Fig.~\ref{fig:cartoon}). 

While  the average molecular orientation is locked in by cross-links beneath the fibril surface, we assume that molecular stretching and compression maintains a global D-band during fibril growth. This local strain is consistent with the flexibility of individual collagen molecules, which have a persistence length of one third of the fully extended molecular length.\cite{Rezaei:2018} Our assumption of a global D-band follows previous theoretical treatment of D-banded fibrils,\cite{Cameron:2020} and from the experimental observation of a robust, well-defined, global D-band for a wide variety of fibril radii.\cite{Baldwin:2020, Berenguer:2014, Hodge:1989, Quan:2015} This assumption also considerably simplifies our approach.

At the fibril surface, we consider two energetic contributions: the Frank free energy density $\tilde{f}_\text{Frank}$, which describes the energetic cost of distortions in the director field, and a phase-field crystal-based free energy density of the D-band, $\tilde{f}_\text{D-band}$. There is also a surface energy term that affects growth, but which does not affect the director orientation or the D-band structure.  

We characterize the average local molecular tilt with respect to fibril axis with a `twist' angle $\psi(\tilde{r})$ (see Fig.~\ref{fig:cartoon}). This tilt depends upon the radial distance $\tilde{r}$, but does not depend on the azimuthal angle. This is called a double-twist configuration,\cite{Brown:2014} and in cylindrical coordinates the local director field is $\boldsymbol{n} = -\sin\psi(\tilde{r}) \boldsymbol{\hat{\phi}} + \cos\psi(\tilde{r}) \boldsymbol{\hat{z}}$. Then the Frank free energy density is simply:\cite{Brown:2014}
\begin{equation}\begin{aligned}\label{frankfe}
    \tilde{f}_\text{Frank} & = \frac{1}{2}\tilde{K}_{22}\left( q - \psi' - \frac{\sin(2\psi)}{2\tilde{r}}\right)^2 + \frac{1}{2}\tilde{K}_{33} \frac{\sin^4(\psi)}{\tilde{r}^2}\\
    & - \frac{1}{2}(\tilde{K}_{22}+\tilde{k}_{24})\frac{1}{\tilde{r}}\frac{d}{d\tilde{r}}\sin^2(\psi),
\end{aligned}\end{equation}
where $\tilde{K}_{22}$,$\tilde{K}_{33}$, and $\tilde{k}_{24}$ are the Frank elastic constants representing the deformations of twist, bend, and saddle-splay, respectively.  The $'$ indicates a derivative, so $\psi'= d \psi/d \tilde{r}$. Note that $q$ is a characteristic preferred twist-angle gradient in a cholesteric phase of collagen molecules, and would be inversely proportional to the cholesteric pitch $P=2\pi/q$. 

For a single-mode approximation of the axial D-band molecular density-modulation, $\delta \rho = \tilde{\delta} \cos{\tilde{\eta} \tilde{z}}$, the phase-field crystal free energy density averaged over one D-band period is:\cite{Cameron:2020}
\begin{equation}
\begin{aligned}
    \tilde{f}_\text{D-band} & = \frac{\tilde{\Lambda}\tilde{\delta}^2}{4}\left(\frac{4\pi^2}{d_\parallel^2} - \tilde{\eta}^2\cos^2\psi\right)^2 + \frac{\tilde{\omega}\tilde{\delta}^2}{2}\left(\frac{3}{4}\tilde{\delta}^2 - \chi^2\right),
\end{aligned} \label{dband}
\end{equation}
where $\tilde{\delta}$ is the D-band amplitude, $\tilde{\eta}$ is the D-band wavenumber, $d_\parallel$ is the preferred D-band spacing, and $\chi$ fixes the preferred D-band amplitude. The parameter $\tilde{\omega}$ represents the strength of the D-band double-well potential, and leads to a non-zero D-band amplitude. The $\tilde{\omega}$ term, like the last term in equation~\ref{frankfe}, is typically negative and so contributes to the thermodynamic stability of collagen fibrils. The parameter $\tilde{\Lambda}$ represents the stiffness of the D-band with respect to its spacing, and energetically couples the D-band and double-twist structures. 

Because we have averaged over the axial D-band modulation to obtain equation~\ref{dband}, we have no axial dependence in our free-energy. Because we have also assumed cylindrical symmetry (i.e. our tilt angle $\psi(r)$ is independent of the azimuthal angle in Fig.~1), we only need to consider the radial coordinate $r$ in our free-energy.

We can collect the free-energy contributions per unit length when a new layer of width $\Delta r$ is added to the fibril from $r$ to $r+\Delta r$:
\begin{equation}\begin{aligned}
     F  = 2\pi\int_{r}^{r+\Delta r} r  f_\text{Frank}dr  + 2\pi \int_0^{r+\Delta r} r f_\text{D-band}(\delta,\eta) dr,
\end{aligned}\end{equation}
where we have included the entire D-band free energy as it is globally minimized. Here we have started to use the dimensionless variables $r \equiv q\tilde{r}/(1-k_{24})$, $f \equiv \tilde{f}(1-k_{24})/\left[\tilde{K}_{22}q^2(1+k_{24})\right]$, and $k_{24} \equiv \tilde{k}_{24}/\tilde{K}_{22}$. Note that $|\tilde{k}_{24}| < \tilde{K}_{22}$,\cite{Ericksen:1991} so we have  $-1 < k_{24} < 1$.

We solve for the values of the twist angle $\psi$, the global D-band amplitude $\delta$, and the global D-band wavenumber $\eta$ that minimize $F$ for all $r\geq0$ in the continuum limit $\Delta r\to0$. We do this by solving the Euler-Lagrange equation for each of the three functions $\psi(r)$, $\delta(r)$, and $\eta(r)$. We obtain an implicit equation for the twist angle function $\psi(r)$ at the current fibril radius $r$:
\begin{equation}\begin{aligned}\label{psieq}
    & r - \frac{1}{2}\sin(2\psi) - K\tan(2\psi)\sin^2\psi \\
    & = \Lambda\delta^2\eta^2r^2\left(4\pi^2-\eta^2\cos^2\psi\right)\tan(2\psi).
\end{aligned}\end{equation}
The double-twist equation depends on the global D-band amplitude and wavenumber. They in turn are explicitly given by 
\begin{eqnarray}
    \label{deltaeq}
    \delta^2(r)  & = & 1-\frac{\Lambda}{\omega r^2}\int_0^{r}x\left(4\pi^2-\eta^2(r)\cos^2\psi(x)\right)^2dx, \\
    \label{etaeq}
    \eta^2(r) & = & 4\pi^2\frac{ \int_0^{r}x\cos^2\psi(x)dx}{\int_0^{r}x\cos^4\psi(x)dx}.
\end{eqnarray}
These equations depend on three key dimensionless parameters 
\begin{eqnarray}
    K &\equiv& \frac{\tilde{K}_{33}}{\tilde{K}_{22}(1-k_{24}^2)}, \label{K} \\ 
    \Lambda & \equiv& \frac{\tilde{\Lambda}\chi^2 (1-k_{24})}{ (1+k_{24})(3d_\parallel^4\tilde{K}_{22}q^2)}, \label{Lambda}\\
    \omega& \equiv&  \frac{\tilde{\omega}\chi^4 (1-k_{24})}{3(1+k_{24}) \tilde{K}_{22}q^2}, \label{omega}
\end{eqnarray}
and also use the dimensionless amplitude
$\delta \equiv \sqrt{3/2}\tilde{\delta}/\chi$ and wavenumber $\eta \equiv \tilde{\eta}d_\parallel$.

The above equations are valid at each fibril radius $r$ during growth. At a current radius $R$, the  double-twist function $\psi(r)$ for $r \leq R$ represents the twist formed when the fibril was at radius $r$, while $\eta(R)$ and $\delta(R)$ characterize the current global D-band. As we will discuss in Section~\ref{sec:radiuscontrol}, what will determine the final fibril radius is the free-energy contribution per unit volume at the fibril surface,
\begin{equation}\begin{aligned}\label{f}
    f(R) & = \frac{1}{2}K\frac{\sin^4\psi(R)}{R^2} + \frac{1}{2}\left(\frac{\sin(2\psi(R))}{2R}\right)^2 - \frac{\sin(2\psi(R))}{2R} +\frac{\gamma}{R}\\
    & + \omega\left[ \delta^2 \left(\frac{\delta^2}{2}-1\right) + R\delta\delta'\left(\delta^2-1\right)\right]\\
    & + \Lambda \left[ \frac{\delta^2}{2}\left(4\pi^2 - \eta^2\cos^2\psi(R)\right)^2 \right.\\
    & \; \; \; \left. + 4\pi^2\delta\delta'\left(2\pi^2 R - \eta^2\frac{1}{R}\int_0^R r\cos^2\psi(r) dr\right)\right],
\end{aligned}\end{equation}
where we have used the current D-band $\eta(R)$ and $\delta(R)$. Here we introduce the dimensionless surface tension $\gamma \equiv \tilde{\gamma}/\big(\tilde{K}_{22}q \big(1+k_{24}\big)\big)$ where $\tilde{\gamma}$ characterizes the surface tension between the fibril surface and the surrounding solution.  $\gamma$ affects fibril structure only through fibril growth (see below). Note that $f=0$ for a bulk cholesteric phase.

We numerically solve equations~\ref{psieq}-\ref{etaeq} using Newton's method. Starting from a numerical solution to equation~\ref{psieq} with $\Lambda =0$, we alternately compute the D-band variables based on the  twist function, and then a new twist function based on the D-band variables. This process is iterated until convergence is achieved. We then numerically compute the free energy density, equation~\ref{f}. The code used for our numerics is available on GitHub.\cite{github} 

\begin{table}[htb]
\centering
\begin{tabular}{|c l l|} 
\hline
 $\psi$ & Twist angle  &Fig.~\ref{fig:cartoon}\\ 
 $r$ & Radial coordinate & Fig.~\ref{fig:cartoon}\\
 $R$ & Fibril radius  &\\
 $\delta$ & D-band amplitude &Eqn.~\ref{deltaeq}\\
 $\eta$ & D-band wavenumber &Eqn.~\ref{etaeq}\\
 $K$ &  Frank coupling &Eqn.~\ref{K}\\ 
 $\Lambda$ &  D-band coupling&Eqn.~\ref{Lambda}\\ 
 $\omega$ & D-band energy &Eqn.~\ref{omega}\\ 
  $f$ & Free-energy density &Eqn.~\ref{f}\\
 $\gamma$ &  Surface tension & \\ 
 \hline
\end{tabular}
\caption{Key dimensionless variables and parameters, and where they are defined.}
\label{table:paramtable}
\end{table}

While we have introduced a flurry of dimensionless quantities (summarized in Table~\ref{table:paramtable}), they significantly simplify our results. In the appendix we discuss various estimates, from which we obtain $K\gtrsim 1$, $\Lambda \geq 0$, $\gamma \in [0.005, 0.3]$, and $\omega \in [0, 20]$. These estimates provide us with some guidance in choosing parameters for our numerical studies.

We will confront our model results with experimental observations on collagen fibrils from tendon and corneal tissues. Tendon fibrils can be characterized by a broad distribution of large radius fibrils (50nm to 200nm), and a small surface twist (no more than $5^\circ$) \cite{Reale:1981}. Corneal fibrils on the other hand are defined by small, tightly distributed, radii (approximately 15nm) and very high surface twist ($\approx16^\circ$).\cite{Raspanti:2018, Holmes:2001} Both types of fibrils show the axial D-banding modulation that is found ubiquitously in animals from insects \cite{Gray:1959} to Dinosaurs \cite{Boatman:2019} with  period between 64 and 67 nm.

\begin{figure*}[h!] 
\centering
  \includegraphics[width=17.1cm]{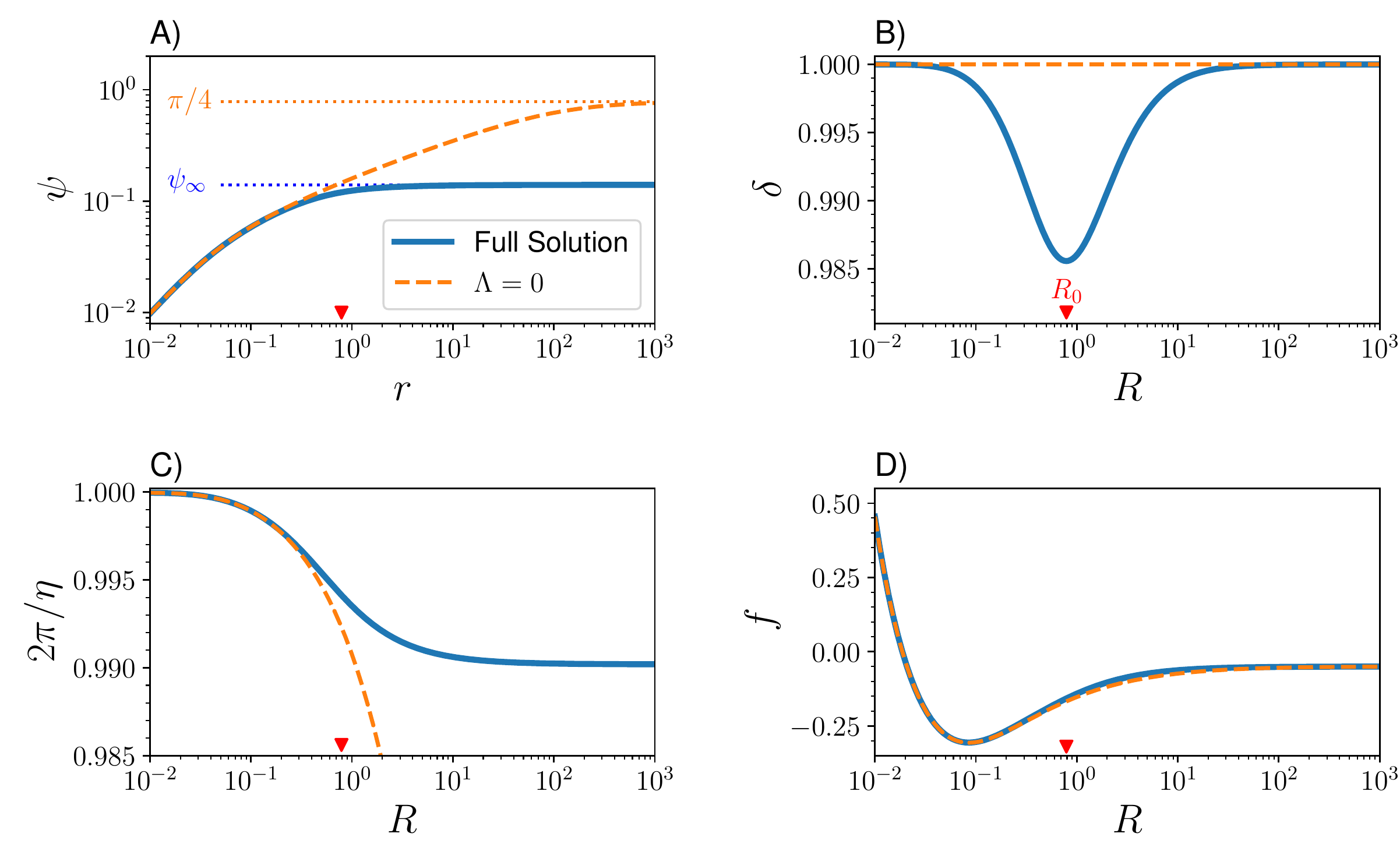}
\caption{Typical non-equilibrium fibril structure. Using the numerical solution to equations~\ref{psieq} to~\ref{f}, we show how \textbf{A)} the twist angle $\psi$, \textbf{B)} the D-band amplitude $\delta$, \textbf{C)} the D-band period $2\pi/\eta$, and \textbf{D)} the surface free-energy density $f$ vary with radius during growth. $\psi(r)$ indicates the twist-function within a fibril (for any  radius $r \leq R$, where $R$ is the fibril radius); $\delta(R)$ and $\eta(R)$ represent a global D-band value for fibrils of radius $R$; while $f(R)$ represents the free-energy density at the surface of a fibril of radius $R$. The red triangles indicate the dimensionless radius $R_0$ where $\delta$ is minimized. With dashed orange lines, we also show results for fibrils in the limit with no coupling between the D-band and double twist ($\Lambda \rightarrow 0^+$). In \textbf{A)} we indicate the limiting values for the twist angle $\psi_\infty$ for both the general solution (blue dotted lines) and the $\Lambda=0$ solution (orange dotted). Parameters used are $K=100$, $\Lambda=0.5$ (for blue lines) or $\Lambda=0^+$ (for orange lines), $\omega=0.1$, and $\gamma = 0.01$. All parameters and values are dimensionless -- as defined in the text.}
  \label{fig:solution}
\end{figure*}

\section{Results}
When $\Lambda=0$, without any coupling between the D-band and the twist function, from equation~\ref{psieq} we find that $\psi \sim r$ at small $r$ while as $r \rightarrow \infty$ the twist-angle increases towards $\psi_{\infty}=\pi/4$. For $\Lambda >0$ a qualitatively similar behavior is observed, though with $\psi_{\infty}< \pi/4$.  A power series expansion in $r$ is possible, see Appendix~\ref{sec:Limiting}. Linearity at small $r$ is maintained because the $\Lambda$ dependence on the right-side of equation~\ref{psieq} enters at $O(r^2)$ -- i.e. it vanishes at small $r$. By the same token, the right-side dominates when $r$ is large and forces $\psi$ to approach a constant value.

\subsection{Structure}
The thicker blue lines in Figs.~\ref{fig:solution}A-D illustrate the solution to our non-equilibrium fibril structure; the results are qualitatively independent of parameterization. In Fig.~\ref{fig:solution}A we confirm that $\psi$ approaches a constant ($\psi_{\infty}$) for large $r$.  At smaller radii, the twist angle increases approximately linearly -- then it crosses over to a constant $\psi_\infty$ around a characteristic radius $R_0$ (indicated by the red triangle).

For both small and large fibril radii $R$, the D-band amplitude  $\delta \approx 1$ as indicated in Fig.~\ref{fig:solution}B. This confirms our understanding that the D-band is independent of the twist function at small $r$, and is dominant at large $r$ -- and so can be satisfied in both extremes. Between these limits the amplitude exhibits a minimum;  we define $R_0$ to be the location of this minimum of $\delta(R)$.

The D-band period $2\pi/\eta$ monotonically decreases with increasing radius $R$ -- as illustrated in Fig.~\ref{fig:solution}C.  The monotonicity of $\eta$ follows from the observed monotonicity of the twist angle $\psi$ as can be confirmed by taking the derivative of equation~\ref{etaeq} with respect to radius. The monotonicity of $\psi$ is always observed for the parameter ranges explored in this paper, but has not been proven to hold generally.  We predict that larger fibrils have shorter D-band spacing compared to smaller fibrils -- when grown under the same conditions (e.g. within the same tissue). 

Finally, the free energy per unit volume of added fibril $f$ depends strongly on the current fibril radius $R$ -- as shown in Fig.~\ref{fig:solution}D.  We find that, as long as the dimensionless surface tension $\gamma$ is not too large (see appendix Fig.~\ref{fig:gamma}), there exists a global minimum of $f$.  For smaller values of surface tension $\gamma$, the minimum is always at a smaller radius than $R_0$. 

In Figs.~\ref{fig:solution}A and D the dashed orange lines indicate the solution without D-band coupling, with $\Lambda=0$. The twist angle function $\psi$ is roughly the same for small $r \lesssim R_0$, but approaches  $\psi_\infty = \pi/4$ for large radii. This confirms that a smaller surface twist, with $\psi_\infty<\pi/4$, is the result of the D-band coupling. Interestingly, the free energy landscape is only very slightly changed (near $R_0$) by the removal of the D-band coupling. This is because we have kept the uncoupled D-band energetics, with $\omega>0$. 

$R_0$, the crossover radius at which the D-band interaction begins to affect the fibril structure, depends on both $K$ and $\Lambda$ (see appendix Fig.~\ref{fig:r0location}). For either larger $\Lambda$ or smaller $K$, $R_0$ is small -- reflecting an earlier transition to D-band dominated structure (with $\psi \approx \psi_\infty$) with radius. 

\begin{figure}[t] 
\centering
  \includegraphics[width=8.3cm]{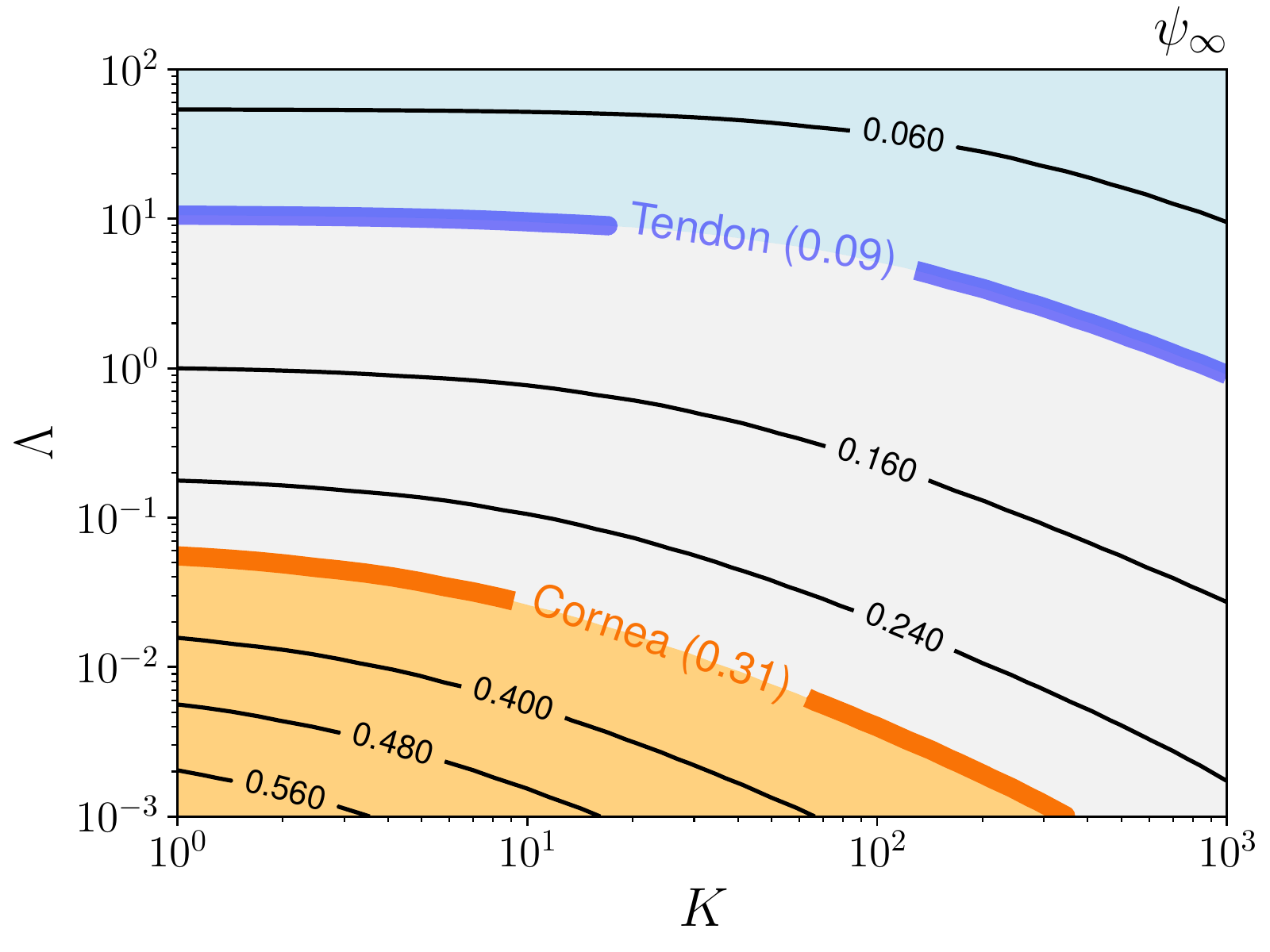}
  \caption{The asymptotic twist angle $\psi_\infty$ for $r \gg R_0$ versus $\Lambda$ and $K$. We show the region where the asymptotic twist is less than or equal to the surface twist of tendon fibrils in blue, and the region where corneal surface twist can be obtained for finite radii in orange. Parameters used are the same as in Fig.~\ref{fig:solution}.}
  \label{fig:asymptotictwist}
\end{figure}

$\psi_\infty$, the asymptotic twist angle observed for  $r \gg R_0$, is shown in Fig.~\ref{fig:asymptotictwist} for various values of $K$ and $\Lambda$. (We find that $\psi_\infty$ is mostly independent of $\omega$, as shown in the appendix Fig.~\ref{fig:omegadependence}.) One contour of interest (orange line) corresponds to the typical reported surface twists of corneal fibrils ($\simeq 17^\circ$ or $0.31$ radians)\cite{Yamamoto:2000, Holmes:2001, Hirsch:2001, Raspanti:2018}  Since the surface twist for fibrils of a given finite $R$ will always be less than $\psi_\infty$, corneal fibrils should be somewhere in the orange-shaded lower-left corner of Fig.~\ref{fig:asymptotictwist} -- below the ``cornea'' line.

We believe that most tendon fibrils are in the $R > R_0$ regime, where $R_0$ is the characteristic core radius. We return to this point in the discussion, but for now we motivate it by the reported systematic differences between small radius and large radius tendon fibrils that has been reported in the literature.\cite{Yamamoto:1997, Quigley:2018} Furthermore, recent study of larger \textit{in vivo} tendon fibrils found no correlation between fibril radius and D-band period (see \cite{Baldwin:2020}, in "Supplemental Information"). This is consistent with larger tendon fibrils being in the $R > R_0$ regime, since in that regime the global D-band period is independent of radius. Interestingly, this also implies that the surface twist of tendon fibrils is independent of radius over a wide-range of fibril radii. Since the reported surface twists of tendon fibrils are typically upper bounds ($\leq 5^\circ$ or $0.09$ radians)\cite{Baselt:1993, Hulmes:1995, Raspanti:2018} this suggests that tendon fibrils will be somewhere in the blue-shaded upper region of the figure -- at or above the ``tendon'' line.    

\begin{figure}[t!] 
\centering
  \includegraphics[width=8.3cm]{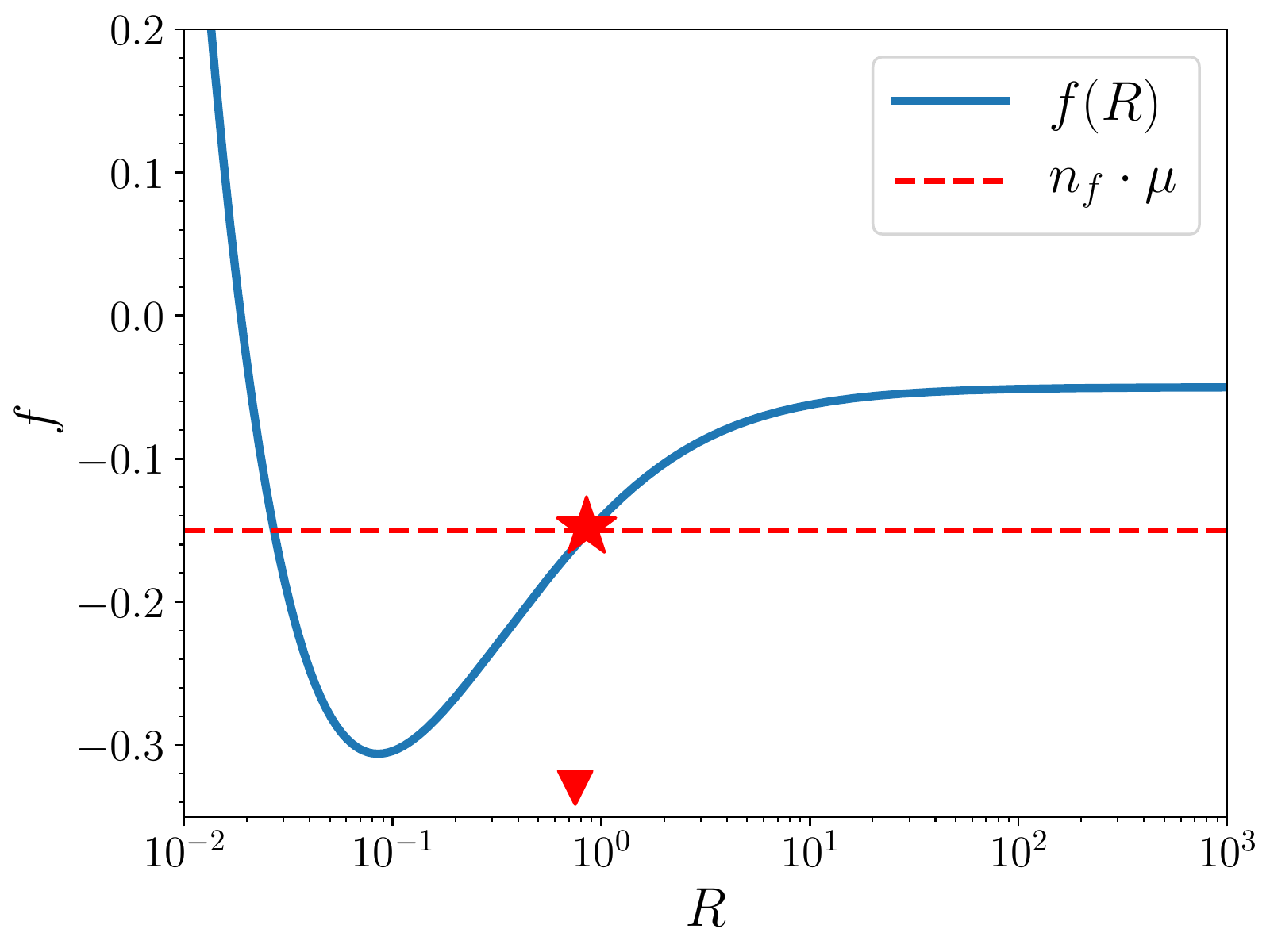}
  \caption{How chemical potential $\mu$ can control fibril radius. The blue line is the same free-energy density of Fig.~\ref{fig:solution}, while the red dashed line is a chemical potential from equation~\ref{mu} -- times the molecular density inside the fibril $n_f$. Fibrils at radius $R$ will grow significantly only if $f(R)< n_f \mu$. The intersection indicated by the red star corresponds to a stable fixed point $R^\ast(\mu)$. At this value of $\mu$, fibrils with $R \lesssim R^\ast$ will grow, while those with $R > R^\ast$ will not.\cite{ratchet}}
  \label{fig:radius_control}
\end{figure}

\subsection{Radius control}
\label{sec:radiuscontrol}
Unlike equilibrium fibril formation,\cite{Brown:2014, Cameron:2018, Cameron:2020} where a fixed fibril radius is determined by global free-energy minimization, for our non-equilibrium model fibril growth is determined by the diffusive supply of collagen molecules from the medium surrounding collagen fibrils.\cite{Rutenberg:2016} When the chemical potential of the medium exceeds the free-energy per molecule within the fibril then growth continues; growth stops when the chemical potential is smaller.\cite{ratchet} This is the condition imposed by local chemical equilibrium between the fibril surface and the extrafibrillar medium.

In Fig.~\ref{fig:radius_control}, we show (solid blue curve) the free energy density per molecule at the fibril surface $f(R)$. The dashed red line indicates $n_f \mu$, where the dimensionless chemical potential per molecule is $\mu \equiv \tilde{\mu} (1-k_{24})/\left[\tilde{K}_{22}q^2(1+k_{24})\right]$; the factor of $n_f$ gives the free energy per unit volume of the fibril so that we can compare directly with $f$.  Under the assumption that collagen molecules in the medium outside the fibril are dilute, we can approximate $\tilde{\mu}$ with the Sackur-Tetrode equation for an ideal gas, 
\begin{equation} \label{mu}
    \tilde{\mu} = k_BT \ln{n/n_c},
\end{equation}
where the number density in the medium is $n$. We use a reference concentration $n_c \in [40,80]$ mg/ml (depending on conditions \cite{DeSaPeixoto:2011}) that supports a cholesteric phase, so that $\tilde{\mu}=0$ at the cholesteric energy $f=0$. 

Fibril growth occurs when the dimensionless $f$ exceeds $ n_f \mu$, so the red star in Fig.~\ref{fig:radius_control} indicates a stable fixed point of fibril radius control. Here, if the collagen concentration in the surrounding medium is kept constant then all fibrils will achieve the same radius -- independent of when they were nucleated.  However, growth of fibrils would slow considerably as the fixed point solution is approached. Furthermore, variations in the concentration within the collageneous tissue would lead to variations in the radius. With dimensioned units, and using equation~\ref{mu}, we have $\tilde{f}' \sigma_{\tilde{R}}  \approx n_f \Delta \tilde{\mu} \approx n_f k_B T \Delta n/n$ -- where $\sigma_{\tilde{R}}$ is the width of the distribution of fibril radii induced by variations of concentration.  This gives us 
\begin{equation}\label{stable}
\begin{aligned}
     \frac{\sigma_{\tilde{R}}}{\tilde{R}} \approx  \frac{ n_f k_B T}{\tilde{f}' \tilde{R}} \frac{\Delta n}{n}. 
\end{aligned}
\end{equation}

We expect that corneal fibrils, which have tight radius control with $\sigma_{\tilde{R}}/\tilde{R} \simeq 0.16$, \cite{Cox:1970} are described by this fixed point. With appropriate dimensions and parameter bounds we show in the appendix (see Fig.~\ref{fig:deltann}A) what concentration control is needed to obtain the observed $\sigma_{\tilde{R}}/\tilde{R}$ for various $K$ and $\Lambda$. We find that the fractional variability of concentration in the medium must be less than $10\%$, which seems plausible. 

Fibrils with $R \gg R_0$ would need to control concentration very finely to achieve stable fixed-point control of fibril radius with equation~\ref{stable} -- due to the weak dependence of free-energy with radius in that regime. Instead we suggest that the growth of all fibrils in this regime is turned on or off by temporal control of the chemical potential (i.e. the \emph{in vivo} concentration $n(t)$ would vary with time $t$ ). Because growth would thereby occur for all fibrils, this would lead to large variations in fibril radius, as is observed in tendon fibrils.  In particular, fibrils that were formed earlier would be larger -- while later fibrils in the same tissue would be smaller. While crude, this mode of regulation allows rapid fibril growth with $n_f \mu \gtrsim f$ for $R > R_0$. 

\section{Discussion and Conclusion} 
Motivated by the diffusional size-exclusion of the cross-linking enzyme lysyl oxidase (LOX) from the interior of growing collagen fibrils, we have developed a model for non-equilibrium \textit{in vivo} fibril formation in which cross-linking only occurs at the fibril surface during growth and in which the fibril interior does not relax its structure significantly during fibril growth due to buried cross-links. Our model is  illustrated in Fig.~\ref{fig:cartoon} and represented by equations~\ref{psieq}-\ref{etaeq}.

Our model predicts two very different regimes of fibril structure, separated by a characteristic radius $R_0$. For small radius fibrils ($R \lesssim R_0$), the twist angle function is linear with $r$ and there are unstrained D-band density modulations within the fibril. For large radii ($R \gg R_0$) we find an approximately constant molecular twist angle at $r > R_0$ and D-band modulations that are compressed below their equilibrium configuration. Stable radius control is possible in the $R \approx R_0$ regime with a static chemical potential of collagen in the surrounding medium. When $R \gg R_0$, fibril radius can only be controlled by switching the growth of all fibrils on or off through temporal control of the chemical potential.

Our results are  agnostic to the detailed structure and properties of the inter-molecular cross-links. Further, our model is compatible with heterogeneous cross-link distributions within collagen fibrils. We require only that the cross-link number density $\rho$ is  always large enough that the energy scale of cross-linking, $\rho k_BT$,\cite{Warner:1996} is  greater than the energy scale of the Frank free energy, $\tilde{K}_{22}q^2$. We estimate that this requires molar densities of $ \gtrsim 0.03-0.3$ cross-links per collagen molecule (see appendix~\ref{sec:crosslink}).  In comparison, $\gtrsim 1.4$ divalent cross-links per molecule are created by LOX across tissue types \emph{in vivo}.\cite{Saito:1997, Depalle:2015}

Our coarse-grained parameters are estimated in Appendix~\ref{sec:Parameterization}. These parameters will depend on biochemical and biological conditions during fibril formation, such as pH and temperature, molecular details of collagen type,\cite{Ricard-Blum:2011} and the presence of any associated proteins that associate with the fibril surface \cite{Rutenberg:2016}. Other molecular details, such as the formation or arrangement of microfibrils,\cite{Zhou:2016} would also affect our parameters.  Microfibril arrangement appears to differ between pre-mineralized bone and tendon tissues,\cite{Zhou:2016} so we expect different coarse-grained parameters between those tissues.

We have assumed that a global D-band is maintained throughout the fibril during growth. This assumption is supported by transmission electron microscopy observations of a well-defined set of molecular density sub-bands within the D-band pattern over a wide range of fibril radii,\cite{Hodge:1989, Quan:2015} as well as similarly sharp D-band spacing across a tissue in x-ray studies\cite{Berenguer:2014} and across fibrils of different radius in atomic force microscopy (AFM) studies.\cite{Baldwin:2020} 

\subsection{Estimating $R_0$ in tendon}
The broad, over twenty-fold,\cite{Kalson:2015, Goh:2012, PattersonKane:1997, Parry:1978} distribution of radii found within the same tendon tissue can be explained by assuming that most tendon fibrils are in the $R \gg R_0$ regime where stable radius control is not practical.\cite{equilibrium} In this regime, fibrils nucleated earlier could have much larger radii than fibrils nucleated later. This implies that within larger tendon fibrils there is a core for $r \lesssim R_0$ (see Fig.~\ref{fig:solution}A). Furthermore, in our model fibrils transition from $R \lesssim R_0$ linear-twist to $R > R_0$ core-shell structures as they grow -- without any changes of the materials parameters (i.e. $K$, $\Lambda$, $\omega$, and $\gamma$). We therefore could expect potentially different behavior for smaller tendon fibrils, with $R \lesssim R_0$. 

While there have been few systematic studies of the structure of collagen fibrils of different radius $R$, there was a remarkable investigation of tendon-like scleral fibrils using AFM techniques by Yamamoto in 1997.\cite{Yamamoto:1997} They measured the depth of the D-band ($\delta$) vs fibril diameter ($2 \tilde{R}$). For $\tilde{R} \gtrsim 25$nm, they found constant $\delta$, while for $\tilde{R} \in [10,25]$nm they found a reduced $\delta_0 \approx 2/3$ (scaled with respect to the value at larger radii). This is  qualitatively consistent with our results, with $\tilde{R}_0 \approx 25$nm, though with a more dramatic decrease than seen in Fig.~\ref{fig:solution}. As shown in the appendix (Fig.~\ref{fig:omegadependence}C), we can recover the scale of the Yamamoto result at small values of $\omega$ -- where a weaker D-band can be suppressed by a modest fibril twist. We expect that parameters could vary significantly between different tissues, so a smaller $\omega$ may not be representative of tendon fibrils more generally. We suggest that $\tilde{R}_0$ and $\delta_0$ should be directly assessed by similar AFM studies of the D-band amplitude over a variety of fibril radii in tendon tissues.\cite{Kalson:2015, Goh:2012, PattersonKane:1997, Parry:1978}

In tendon fibrils, there is some evidence both for \cite{Gutsmann:2003} and against \cite{Wenger:2008, Strasser:2007} mechanical core-shell structure. However, non-enzymatic cross-linking due to advanced glycation end-products (AGE) forms very slowly and may be localized to fibril surfaces after growth completes.\cite{Slatter:2008} This could lead to mechanical contrast between fibril interior and surface. Enzymatic cross-linking at the final fibril surface could lead to a similar effect. This represents an alternative mechanism for any observed elastic core-shell structure.

Nevertheless, recent experimental measurements of \textit{in vivo} extracted positional bovine tendon fibrils under strain have found that radial structure of collagen fibrils can be separated into two components which behave differently under strain: a core around the center of the fibril, in which the D-band structure is fairly robust to mechanical strain, and a shell at the surface of the fibril, which undergoes significant structural deformation when the fibril is stretched to failure.\cite{Quigley:2018} Furthermore, \citet{Quigley:2018} found that smaller radius fibrils extracted from  energy-storing bovine flexor tendons do not show a core-shell response when stretched to failure. Flexor fibrils are $\approx 40\text{nm}$ in radius.\cite{Herod:2016} 

Given the similarity of radial scales between the flexor tendons and the scleral core, we suggest that $\tilde{R}_0 \in [25,40]nm$ in tendon, and that the $r \lesssim R_0$ ``linear twist'' and the $r > R_0$ ``constant twist'' regimes of our model corresponds to the observed core and shell structures, respectively. The larger pre-compression of the interior of large fibrils, shown in Fig.~\ref{fig:molecularstrain}, may provides a mechanism for the differing damage resistance for the core region. 

D-band amplitude studies for fibrils of different radius should be able to distinguish the two possible mechanisms for core-shell structure: we  expect D-band amplitude effects for our structural model, but they are not expected to arise with cross-linking heterogeneity. More generally, we would expect that surface-localized cross-linking after fibril formation would lead to a constant-width {shell} region of the surface independent of fibril radius. In contrast, in our model the {core} has the same width ($R_0$) independent of fibril radius, while fibrils with $R \lesssim R_0$ would have no shell at all. 

\subsection{Corneal fibrils}
We easily obtain the high surface twist of corneal fibrils (see Fig.~\ref{fig:asymptotictwist}), a result that has eluded previous equilibrium treatment of D-banded fibrils.\cite{Cameron:2020} Furthermore, we are able to achieve the corneal twist together with stable fixed-point control of fibril radius at $R \approx R_0$ -- as tuned by the chemical potential of collagen molecules in the medium outside the fibril. 

Corneal fibrils are systematically larger near the corneal periphery, with up to a two-fold increase of radius between the central and peripheral cornea.\cite{Boote:2003} Such a difference can be accommodated in our model without invoking changes in fibril composition -- it implies systematically larger molecular collagen concentrations in the corneal periphery during fibril formation (see Fig.~\ref{fig:radius_control}). Following Fig.~\ref{fig:solution}, for larger (peripheral) corneal fibrils we also expect slightly larger surface twists, larger or smaller D-band amplitudes ($\delta$) depending on whether $R_C/R_0$ is larger or smaller than $1$ (see appendix Fig.~\ref{fig:deltann}), respectively, and a slightly shorter D-band period ($2 \pi/\eta$). 

Within our model, corneal fibrils must be controlled by different dimensionless constants than tendon fibrils. Our model only exhibits monotonically increasing $\psi(r)$, while narrower corneal fibrils have higher surface twist than wider tendon fibrils. As illustrated in Fig.~\ref{fig:asymptotictwist}, it is possible to go from the tendon to the corneal regime by decreasing $\Lambda$ by approximately two orders of magnitude. This appears to be mediated, at least in part, by the collagen composition in different tissues. Corneal fibrils are mostly comprised of collagen type I, but have $\approx 20\%$ of type V. Interestingly, pure type V fibrils have no visible D-band.\cite{Birk:1990} In our model smaller D-band amplitudes ($\chi$, see equation~\ref{dband}) lead to smaller $\Lambda \equiv \tilde{\Lambda}\chi^2 (1-k_{24})/ \left[(1+k_{24})(3d_\parallel^4\tilde{K}_{22}q^2)\right]$. In qualitative agreement, AFM measurements indicate that corneal D-band amplitudes are approximately $1/3$ that of tendon-like scleral fibrils,\cite{Yamamoto:1997} which would account for one order of magnitude difference in $\Lambda$ between tendon and corneal fibrils. Different values of $k_{24}$ can lead to the same effect. We suggest that an admixture of type V collagen in corneal fibrils effectively reduces $\Lambda$ to reach the ``corneal'' region in Fig.~\ref{fig:asymptotictwist}.\cite{medium}

\subsection{Discriminating between equilibrium and non-equilibrium models}
How does our non-equilibrium model of fibril growth and structure compare to equilibrium models,\cite{Cameron:2020} which are more appropriate for \emph{in vitro} studies? The equilibrium picture determined the fibril radius and structure through global free-energy minimization, and obtained a single fibril radius and double-twist structure for a given parameterization. Small radius linear and large-radius constant double-twist structures were identified as distinct equilibrium phases, though coexistence between the phases could be observed for special parameterizations.\cite{Cameron:2020}

Conversely, the non-equilibrium fibril growth described in this paper is controlled by tuning the chemical potential of the collagen surrounding the fibril with respect to the free-energy density at the fibril surface. At every parameterization (corresponding to experimental conditions in a given tissue), fibrils progress from small radius linear to large-radius constant double-twist structures as growth continues. Growth is controlled by either fine-tuning the extrafibrillar collagen concentration to determine a stable-radius with $R \approx R_0$ (see Fig.~4), or by increasing/reducing the extrafibrillar collagen concentration to start/stop growth for all fibrils with $R > R_0$. Because fibril growth will be faster with larger extrafibrillar concentrations, growth can be rapid for $R > R_0$. Rapid growth will not possible for stable radius control with $R \approx R_0$ because the extrafibrillar concentration must be tightly controlled close to a fixed-point.  We believe the $R>R_0$ regime is applicable to tendon fibrils, and the $R \approx R_0$ regime to corneal fibrils. 

Our model presumes that cross-linking takes place during fibril growth at the fibril surface. The relative ease that our non-equilibrium model has in reproducing high surface twists, as compared with an equilibrium approach,\cite{Cameron:2020} suggests that previous difficulties in reconstituting shorter D-band periods associated with highly twisted fibrils\cite{Raspanti:2018, Brodsky:1980} may be more successful if chemical cross-linking agents\cite{JorgeHerrero:1999} were used during fibril assembly \emph{in vitro}. This approach may also be useful to test our model results in controlled conditions.

Conversely, inhibiting cross-linking during \emph{in vivo} fibril formation would also be interesting. We would expect that complete inhibition would recover earlier equilibrium predictions \cite{Cameron:2020}, while weak inhibition would satisfy our assumptions of strong cross-linking with respect to Frank energies. 

Scanning electron-microscopy (SEM) techniques for imaging fibrils have been used to measure the molecular twist angle at the {surface} of collagen fibrils. Determining the twist field {inside} fibrils would provide significantly more information. Transmission electron-microscopy (TEM) studies are laborious and need replication, but can obtain twist fields of single fibrils.\cite{Holmes:2001}  A promising high-throughput technique is polarization-resolved second harmonic generation (P-SHG) microscopy, which uses non-linear optical effects to measure volume-averaged anisotropy. P-SHG anisotropy measurements of collagen fibrils have been made in recent years.\cite{Gusachenko:2012, Rouede:2020} We hope they can be adapted to measure, e.g. the volume-average twist field $\langle \psi(r) \rangle$, which could provide insight into whether individual fibrils are in the $R > R_0$ or $R \lesssim R_0$ regime. 

\subsection{Conclusion} 
Equilibrium coarse-grained models of collagen fibrils have provided a quantitative description of the structure of fibrils that form \textit{in vitro} in the absence of cross-linking enzymes.\cite{Cameron:2020} The presence of LOX during \textit{in vivo} fibril formation requires a different, nonequilibrium, model for fibril growth. In this work we have outlined a simple coarse-grained approach for nonequilibrium fibril growth and structure. We assume only local equilibrium at the surface of growing fibrils, together with no molecular reorientation in the interior of growing fibrils due to cross-linking. The size-exclusion of LOX from the fibril interior indicates the importance of rapid LOX activity at the surface of growing fibrils \emph{in vivo}, including the potential localization of LOX co-factors to the fibril surface.\cite{Rutenberg:2016}

\appendix

\section{Appendix: Model Parameterization}
\label{sec:Parameterization}
In this appendix we use a variety of experimental studies of collagen systems, and collagen-like systems, to get parameter bounds and estimates. Since tendon and corneal fibrils provide contrasting structure and are well characterized, constraints from those systems are also used. We know that parameters estimated \emph{in vitro} will not exactly match those seen \emph{in vivo}, nor parameters estimated in dense cholesteric solutions vs. the fibril interior, nor parameters estimated in largely collagen type I fibrils vs. parameters for other fibrillar collagens. These are order of magnitude estimates only. Nevertheless, we hope they outline what is known, what can be studied, and what would be important to better characterize. 

\begin{figure*}[tb!] 
   \centering
   \includegraphics[width=17.1cm]{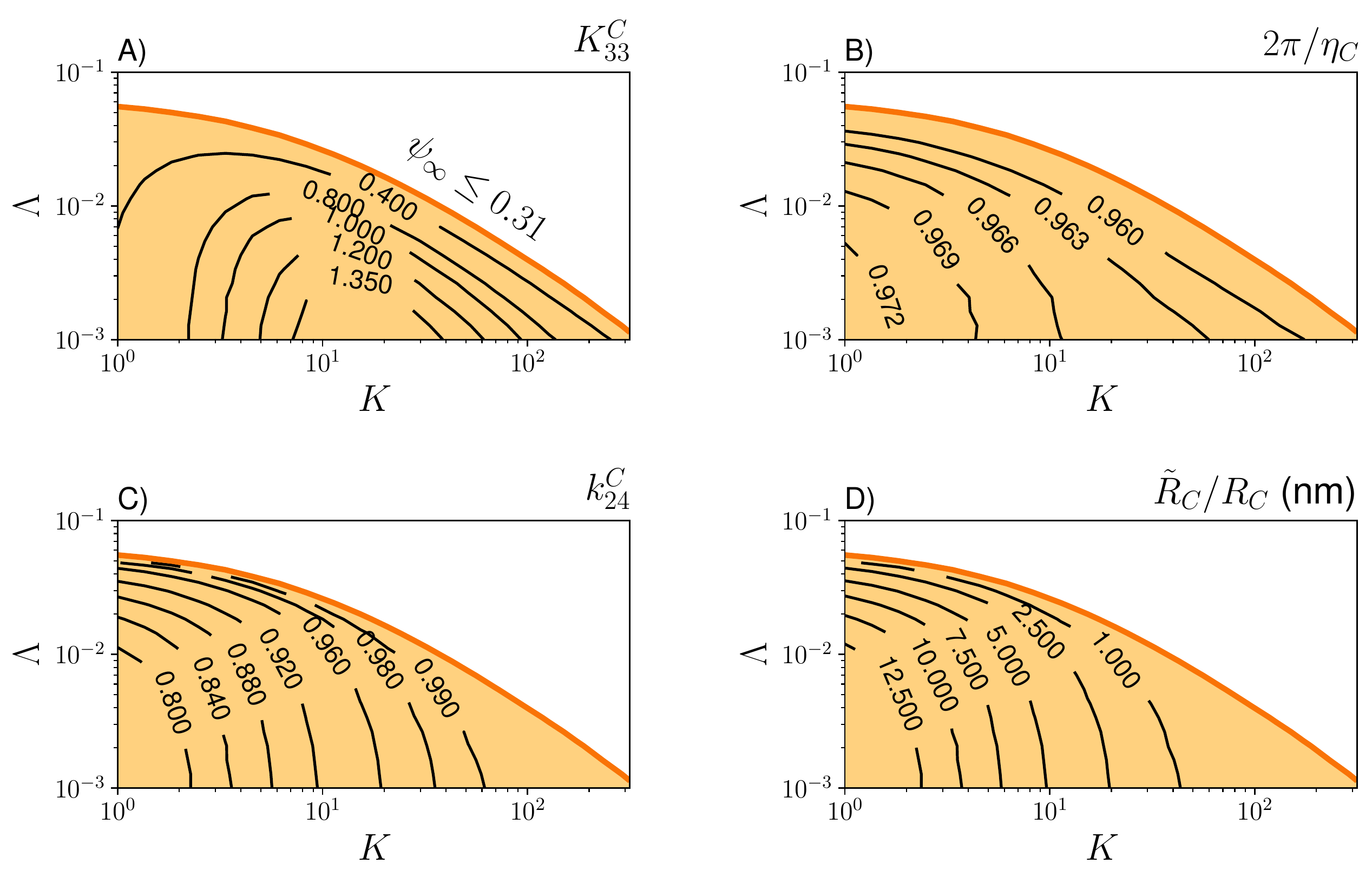} 
   \caption{Possible corneal parameterizations for a variety of $K$ and $\Lambda$, using a dimensionless fibril radius $R_C$ such that $\psi(R_C) = 0.31$. This restricts us to the orange shaded region where $\psi_\infty \leq 0.31$. A) $K_{33}^C$, using equation~\ref{K33}. B) D-band period $2 \pi/\eta_C$. This is relative to the period observed for untwisted fibrils, so we expect $64nm/67nm \approx 0.96$.\cite{Raspanti:2018} C) Estimated value of $k_{24}^C$. This is obtained using $k_{24} = 1-q \tilde{R}/R$.  D) The lengthscale $X \equiv \tilde{R_C}/R_C$ (in nm) used to convert between dimensioned and dimensionless units.  We have used $q = 5 \pi \mu m^{-1}$ and a dimensioned radius $R_C = 15nm$.}
   \label{fig:cornea}
\end{figure*}

\subsection{Franck constants}
The pitch $P=2\pi/q$ for dense collagen phases monotonically decreases with concentration in qualitative studies that went up to 1000 mg/ml.\cite{Mosser:2006} The relationship between pitch and concentration was characterized between 50 -400 mg/ml.\cite{DeSaPeixoto:2011} Extrapolating those results to the wet density of tendon tissue, 1000mg/ml,\cite{Ker:1981} leads to a half-pitch of $P\approx 0.4\mu$m. Accordingly, we estimate $q \approx 5 \pi \mu m^{-1}$.

From the definition of $K \equiv K_{33}/(1-k_{24}^2)$ and our dimensionless transformation $R=q \tilde{R}/(1-k_{24})$, we can eliminate $k_{24}$ and obtain a convenient expression for $K_{33}\equiv \tilde{K}_{33}/\tilde{K}_{22}$:
\begin{equation}
    K_{33} = K \left[ 1 - \left(1 - \frac{q \tilde{R}}{R} \right)^2 \right].
    \label{K33}
\end{equation}
Using a dimensioned radius of corneal fibrils $\tilde{R}_C \approx 15$nm and a dimensionless $R_C$ that corresponds to a surface-twist $\psi(R_C) \approx 0.31$ radians, we use our $q$ estimate to show $K_{33}^C$
in Fig.~\ref{fig:cornea}A for various $K$ and $\Lambda$. We see that $K_{33}^C \approx 1$. Theoretical treatments of liquid crystals composed of long rod-like molecules indicate that $K_{33}>1$.\cite{Strayley:1973, Ferrarini:2010} Accordingly, we expect corneal fibrils to occupy the $K_{33}>1$ region of Fig.~\ref{fig:cornea}C.  When  $K_{33} \ll K$, we can expand equation~\ref{K33} to obtain $K_{33} = 2q K \tilde{R}/R$, and we see that our estimated $K_{33}$ is proportional to our assumed $q$. If we have underestimated the effective $q$ in collagen fibrils, then a wider region of $K_{33}>1$ would be available.  

Using the same estimation of $R_C$ that recovers the corneal surface twist $\psi(R_C)$, we show the D-band period $2 \pi/\eta_C$ as a function of $K$ and $\Lambda$ in Fig.~\ref{fig:cornea}B. We recall that $\eta_C$ is with respect to an untwisted fibril, as in Fig.~\ref{fig:solution}A. The corresponding experimental value is the ratio of the corneal to tendon D-band periods, which is $\simeq 64 nm/67nm= 0.96$.\cite{Raspanti:2018} Our results are consistent with this.

We show our estimate of $k_{24}^C = 1 -q \tilde{R}_C/R_C$ in Fig.~\ref{fig:cornea}C. It satisfies the strict requirement that $|\tilde{k}_{24}| < \tilde{K}_{22}$ for nematic liquid crystal systems, so that $-1 < k_{24} < 1$.\cite{Ericksen:1991} We see that $k_{24}^C \in[0.8,1]$ in the region corresponding to corneal fibrils.

In Fig.~\ref{fig:cornea}D we directly show the ratio $X \equiv \tilde{R}_C/R_C$, which is the dimensioned units of length (in nm) corresponding to our dimensionless length. We see that $X \approx 1-15 nm$ in the region corresponding to corneal fibrils.

For the many other kinds of fibrillar collagen, we are much less constrained experimentally. For tendon fibrils in particular, the asymptotic twist is much less precisely measured -- and in our model it does not constrain larger radii with $R > R_0$ due to the weak twist-dependence there.  However, we have already noted that $\Lambda$ needs to be much larger for tendon fibrils than corneal -- see Fig.~\ref{fig:asymptotictwist}. We know that tendon and corneal fibrils cannot have the same parameterization since tendon fibrils have larger radius but smaller twist than corneal -- and our model only exhibits monotonic twist increasing with radius.

\begin{figure*}[t!] 
\centering
  \includegraphics[width=17.1cm]{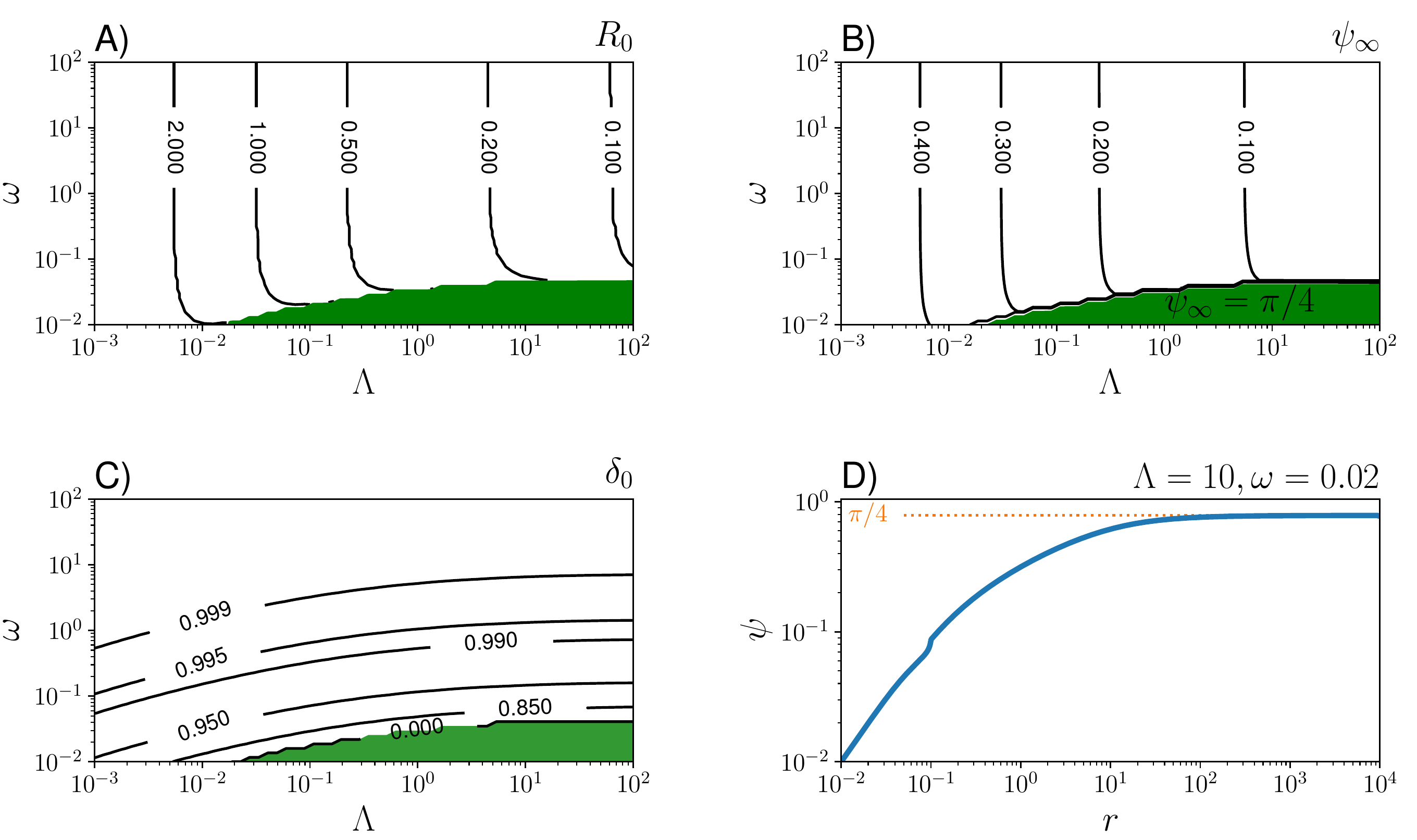}
  \caption{As $\omega$ and $\Lambda$ are varied, with $K=10$. \textbf{A)} the dimensionless radius $R_0$ at which the D-band amplitude is minimized; \textbf{B)} the asymptotic twist angle $\psi_\infty$; and \textbf{C)} the minimum of $\delta_0=\delta(R_0)$. The green shaded region indicates $\delta_0 = 0$ with $\psi_\infty=\pi/4$. \textbf{D)} $\psi(r)$ for $\Lambda=10$, $\omega=0.02$.}
  \label{fig:omegadependence}
\end{figure*}

\subsection{Chemical potential}
We model the extrafibrillar collagen molecules as an ideal gas, and use the Sackur-Tetrode expression (equation~\ref{mu}) to determine the chemical potential. To compare with free energy densities, we want the chemical potential density, or chemical potential per unit volume. This is given by $n_f\tilde{\mu}$, where $n_f$ is the number density of molecules within a fibril. We know the dimensions of collagen molecules ($1.5$nm diameter and $300$nm length) so, given that the packing fraction of molecules within a fibril is approximately 0.7,\cite{Toroian:2007} the number density of molecules within a fibril is  $n_f \simeq 1.32\times 10^{24}$m$^{-3}$. 

Room temperature gives $k_B T=4.28 \times 10^{-21}$J. A lower bound for the number density $n$ of collagen molecules in the cellular environment outside of fibrils is $8$ mg/ml.\cite{Rutenberg:2016} An upper estimate for the critical concentration of extrafibrillar collagen that supports a cholesteric phase is $n_c \approx 80$ mg/ml.\cite{DeSaPeixoto:2011} This  gives $n_f \tilde{\mu} \in [-13,0]$ kPa. We can compare with our dimensionless free energy density by defining a dimensionless chemical potential per molecule,
\begin{equation}
    \mu \equiv \frac{\tilde{\mu}(1-k_{24})}{\tilde{K}_{22}q^2(1+k_{24})}.
\end{equation}
We use $\tilde{K}_{22}\in[0.6,6]$pN.\cite{Cameron:2018} Using also $k_{24} \in [0.8,1]$ and $q \approx 5 \pi \mu m^{-1}$, we obtain a range $n_f \mu \in [-10,0]$. 

\subsection{D-band parameters}
$\omega$ can be bounded above by the polymerization energy of stable collagen fibrils. The change in Gibbs free energy during fibril formation is $\Delta G \simeq -13$ kcal/mol.\cite{Kadler:1987} Using $n_f$, the absolute change in the free energy density due to fibril formation is $|\Delta\langle \tilde{f}\rangle| \approx 120$kPa. Since the cholesteric phase has $f=0$, the free-energy of formation is at least $|f(R)|$ for $f<0$. Since $f = - \omega/2$ as $R \rightarrow \infty$, $\omega/2$ then provides a lower bound of the formation free-energy. Using the dimensionless units of $f \equiv \tilde{f}(1-k_{24})/\left[\tilde{K}_{22}q^2(1+k_{24})\right]$, $q$, the lower end of the estimated range $\tilde{K}_{22} \in [0.6,6]pN$,\cite{Cameron:2018} and $k_{24} \approx 0.8$ from earlier, we obtain $\omega \in [0,200]$.  However, we can do better.

We also require $n_f \mu < f$ for fibril growth, following Fig.~\ref{fig:radius_control}. Since $f<-\omega/2$ generally holds, we use $n_f \mu \in [-10,0]$ to obtain $\omega \in [0,20]$. This is consistent with, but tighter than, the bounds provided by the polymerization energy.

Fig.~\ref{fig:omegadependence} shows  $R_0$, $\psi_\infty$, and $\delta_0$ versus $\omega$ and $\Lambda$, for constant $K=10$. $\Lambda$ is not constrained by experiment, so we explore a wide range of parameter values for $\Lambda \geq 0$. These values are independent of $\omega$ for larger values of $\omega$.  Only when $\omega \lesssim 0.1$, at the very low end of our estimated range and where the characteristic D-band cohesion is very weak, do we see changes due to $\omega$. At these very small $\omega$ the D-band disappears for larger fibril radii, so that   $\delta_0=0$ and $\psi_\infty= \pi/4$ (indicated by the green shaded region). In Fig.~\ref{fig:omegadependence}D we show and example of a twist angle function $\psi(r)$ in the green $\delta_0=0$ region with $\Lambda=10$ and $\omega=0.02$. The small kink at $r \approx 0.1$ corresponds to the radius at which the D-band amplitude vanishes during fibril growth.

\begin{figure}[t!] 
\centering
  \includegraphics[width=8.3cm]{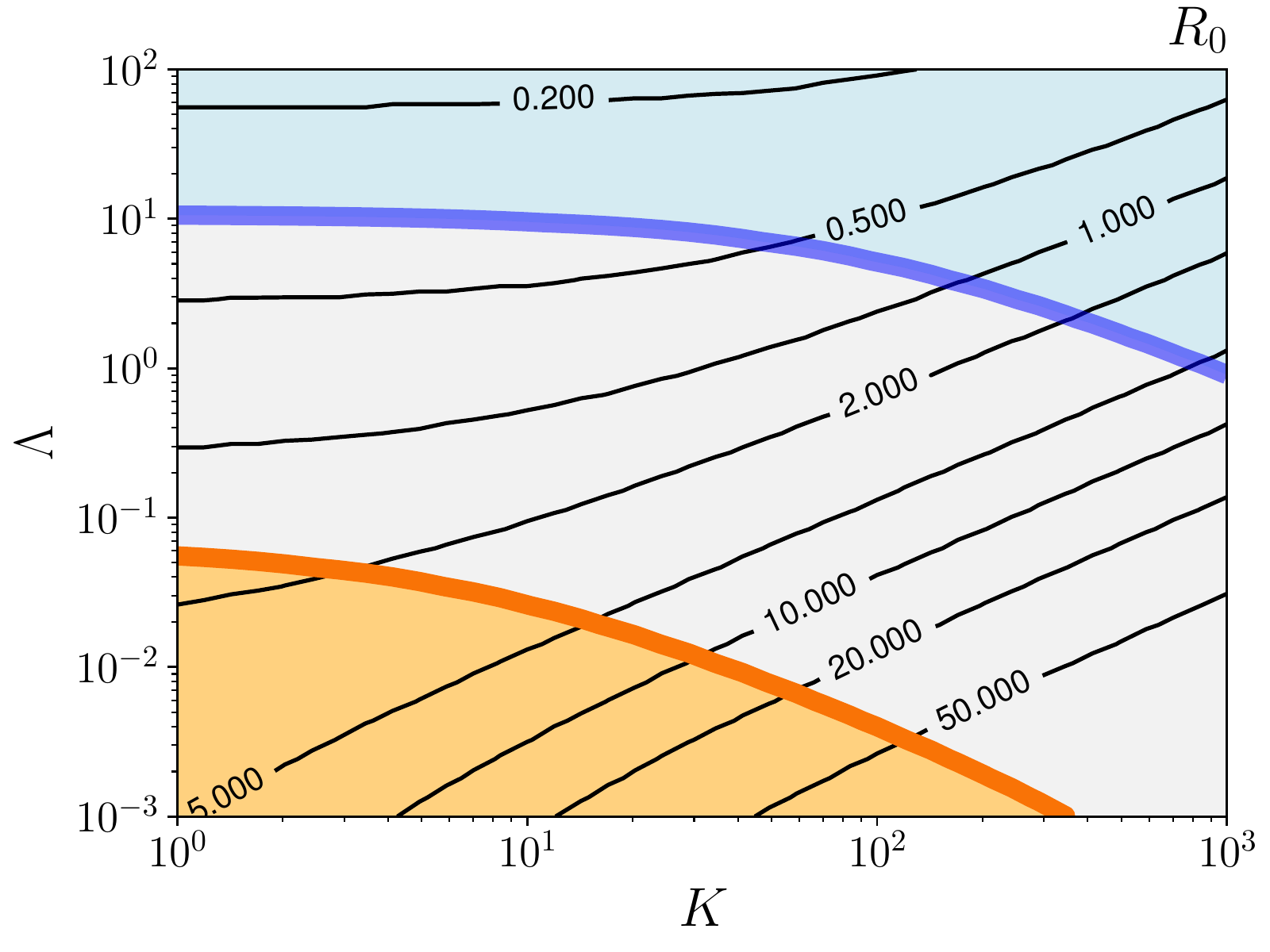}
  \caption{The dimensionless value of $R_0$, the radius at which the D-band amplitude is minimized, as a function of $K$ and $\Lambda$. We use  $\omega=0.1$.}
  \label{fig:r0location}
\end{figure}

\begin{figure}[t!] 
\centering
  \includegraphics[width=8.3cm]{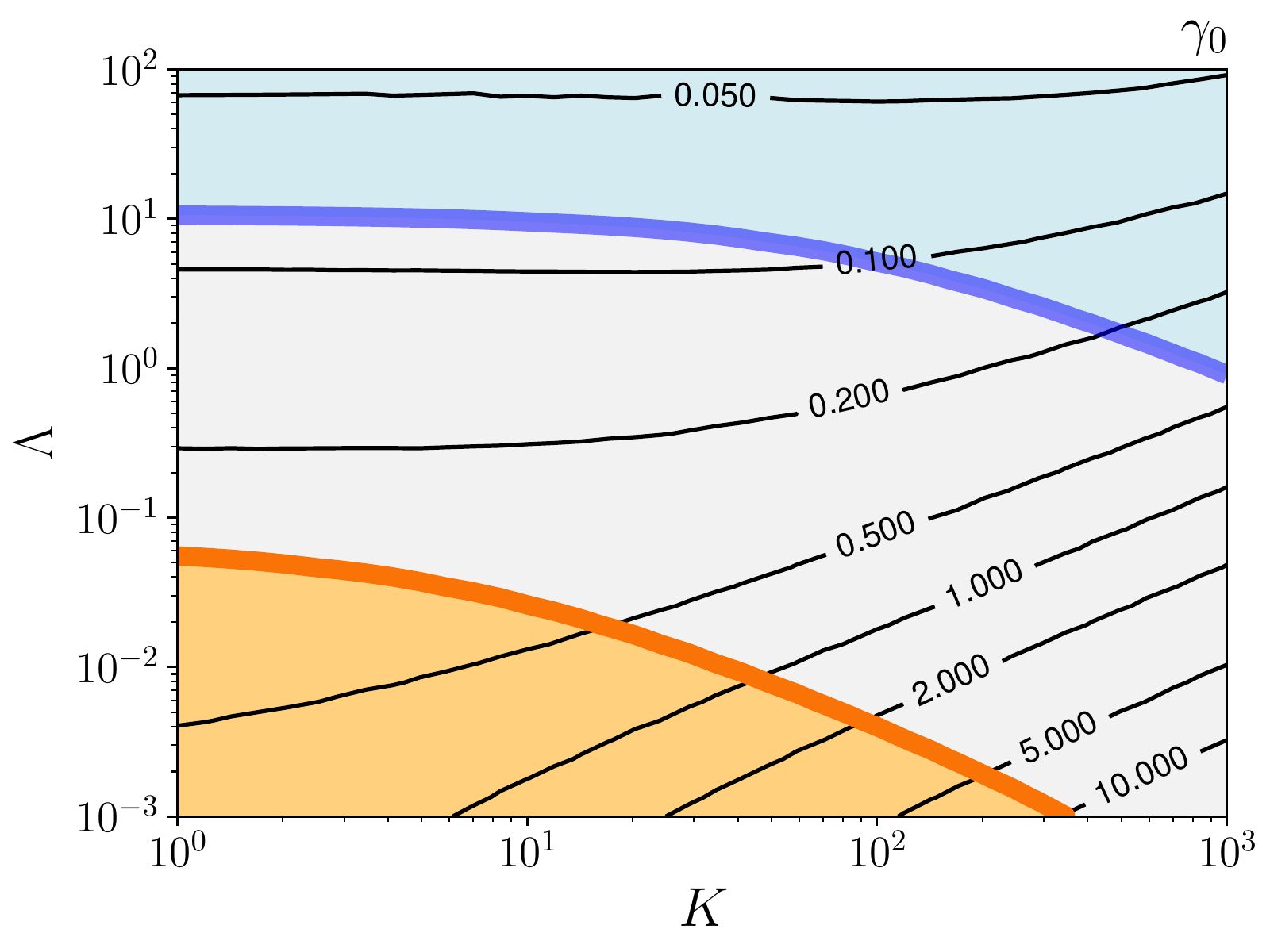}
  \caption{The dimensionless surface tension $\gamma_0$ that is large enough to make $f(R_0)=0$, i.e. to destabilize fibril growth at $R_0$ with respect to a cholesteric phase. In comparison, we estimate that $\gamma \in [0.005,0.3]$.}
  \label{fig:gamma}
\end{figure}

\subsection{$R_0$}
We show the dimensionless $R_0$, the radius at which the D-band amplitude $\delta$ is minimized, vs $K$ and $\Lambda$ in Fig.~\ref{fig:r0location}. We see that we expect  $R_0 \lesssim 5$ for tendon fibrils, while $R_0^C \gtrsim 2$ for corneal fibrils. 

Direct comparison of different fibril types requires the dimensioned $\tilde{R}_0 = (1-k_{24}) R_0/q$. The length unit $X \equiv (1-k_{24})/q$ was determined  as a function of $K$ and $\Lambda$ in Fig.~\ref{fig:cornea}D  -- but only for corneal fibrils. There we observed possible values of $X^C \in [1,20]nm$ for corneal fibrils in Fig.~\ref{fig:cornea}.

We do not know how the length unit $X$ varies for different fibril types as a function of $K$ and $\Lambda$. However, we can explore the hypothesis that tendon fibrils have a core-structure with $\tilde{R}_0 \in [25, 40]nm$ (see discussion). Given $R_0 \in [0,5]$ for tendon, we estimate that $X \gtrsim 5$nm. This overlaps with the range of $X^C$ values. Equivalently, we can constrain $k_{24} < 0.92$ for tendon which overlaps with the range of values of $k_{24}^C$ shown Fig.~\ref{fig:cornea}C. It is possible that $k_{24}$ and $q$ (and hence $X$) are similar in different fibril types. 

\subsection{Surface energy}
Following \citet{Jawerth:2018}, we expect the dimensioned surface energy $\tilde{\gamma}\in [1,5]$pN$\mu$m$^{-1}$. The dimensionless $\gamma = \tilde{\gamma}/\left[\tilde{K}_{22} q (1+k_{24})\right]$. Using $k_{24} \approx 1$, $\tilde{K}_{22} \in [0.6,6]pN$ \cite{Cameron:2018}, and $q$, we expect the dimensionless $\gamma\in [0.005,0.3]$.  Fig.~\ref{fig:gamma} indicates the dimensionless surface tension $\gamma_0$ needed to significantly change the stable radius control picture of Fig.~\ref{fig:radius_control}. 

$\gamma$ does not enter into the structural equations~4-6, but enters directly into $f(R)$ in equation~\ref{f} with $\gamma/R$. At large radius surface tension will have little effect. We estimate the value $\gamma_0$ at which it has a large effect at intermediate radius by equating $\gamma_0/R_0 = |f(R_0)_{\gamma=0}|$. This value of $\gamma_0$ would destabilize fibril growth with respect to a cholesteric phase with $f=0$. From our estimated range of $\gamma$ we see that for tendon fibrils (blue region) we could have a large-enough surface tension to significantly affect $f(R_0)$ -- but that would not affect $f(R)$ for $R \gg R_0$. In contrast, it is unlikely that the dimensionless surface tension is large enough to significantly affect $f(R_0)$ for corneal fibrils.

\subsection{Cross-link energetics}
\label{sec:crosslink}
Our model assumes that there is no relaxation of the double-twist orientation inside the current fibril radius. This amounts to an approximation that the cross-linked elastic fibril is rigid with respect to reorientation. Short of exploring a fully elastomeric model, we can estimate the energy-scales we expect for any such reorientation.  We compare the elastomeric energy scale $\rho k_B T$,\cite{Warner:1996} with the Franck energy scale $\tilde{K}_{22} q^2$. Here $\rho$ is the cross-link number density, and we have $q \approx 5 \pi \mu m^{-1}$, $k_B T=4.28 \times 10^{-21}$J, $\tilde{K}_{22} \in [0.6,6]pN$,\cite{Cameron:2018} and we also use the number density of collagen molecules
$n_f \simeq 1.32\times 10^{24}$m$^{-3}$.

We obtain a lower-bound on the molar concentration of cross-links per collagen molecule of $\rho/n_f \gtrsim \tilde{K}_{22} q^2/\left( n_f k_B T \right) \approx 0.03 - 0.3$. In comparison, experimental assays of cross-link density across tissue types give approximately $1.4$ cross-links per molecule for ``immature'' divalent cross-links \cite{Saito:1997}, and a much more tissue-dependent measurement of at least $0.3$ cross-links per molecule for ``mature'' trivalent cross-links (see \cite{Depalle:2015}). This appears consistent with our assumption that fibrils \emph{in vivo} are strongly cross-linked, and unlikely to significantly relax their orientation after formation.

Nevertheless, we have allowed our D-band period to globally adjust during fibril growth. We justified this phenomenologically, based on the observed sharpness of the D-band seen in experiment.\cite{Baldwin:2020, Berenguer:2014, Hodge:1989, Quan:2015} The energetics of this global adjustment are not included in the $\tilde{\Lambda}$ term of equation~\ref{dband}, so we are assuming that the D-band stiffness $\tilde{\Lambda}$ is significantly above the corresponding elastomeric stiffness. This remains to be investigated.

\section{Appendix: Limiting Behavior}
\label{sec:Limiting}
For small $r$, we can derive a power series solution to equations~\ref{psieq}-\ref{etaeq}. To fifth order, we obtain
\begin{subequations}
\begin{equation}
    \psi(r) \approx r + \left(\frac{2}{3} - 2K\right)r^3 + \left(12K^2 - 10K + \frac{6}{5} - 16\Lambda\pi^4 \right)r^5+ \mathcal{O}(r^7),
\end{equation}
\begin{equation}
    \delta^2(r) \approx 1 - \frac{2\pi^4\Lambda}{3\omega}r^4 + \mathcal{O}(r^6),\text{ and}
\end{equation}
\begin{equation} \label{eta}
    \eta^2(r) \approx 4\pi^2 + 2\pi^2r^2 + \left(2 - \frac{16}{3}K\right) \pi^2 r^4 + \mathcal{O}(r^6).
\end{equation}
\end{subequations}
We can see that the coupling $\Lambda$ of the D-band and Frank energies does not enter the twist function until $\mathcal{O}(r^5)$. Conversely, the Frank parameter $K$ only appears in the  D-band period $\eta$  at $\mathcal{O}(r^4)$ and only in the amplitude $\delta$  at $\mathcal{O}(r^6)$ (not shown). Thus at smaller radius, when the twist-function is approximately linear, the D-band and twist-function are uncoupled. 

For large $r$, an explicit expansion in $1/r$ was not possible since the asymptotic value $\psi_\infty$ is undetermined without a full numerical solution. However, the D-band $\Lambda$ term in equation~\ref{psieq} dominates. Taking a derivative with respect to $r$ confirms that $\psi'\to0$, so that the twist angle approaches a constant value $\psi_\infty$ at large radius. It then follows that as $r \to \infty$ we also have $\eta \to 2\pi/\cos \psi_\infty$, $\delta \to 1$, and $f \to -\omega/2$. 

\section{Appendix:Molecular strain}
\label{sec:MolecularStrain}
\begin{figure}[t!]  
\centering
  \includegraphics[width=8.3cm]{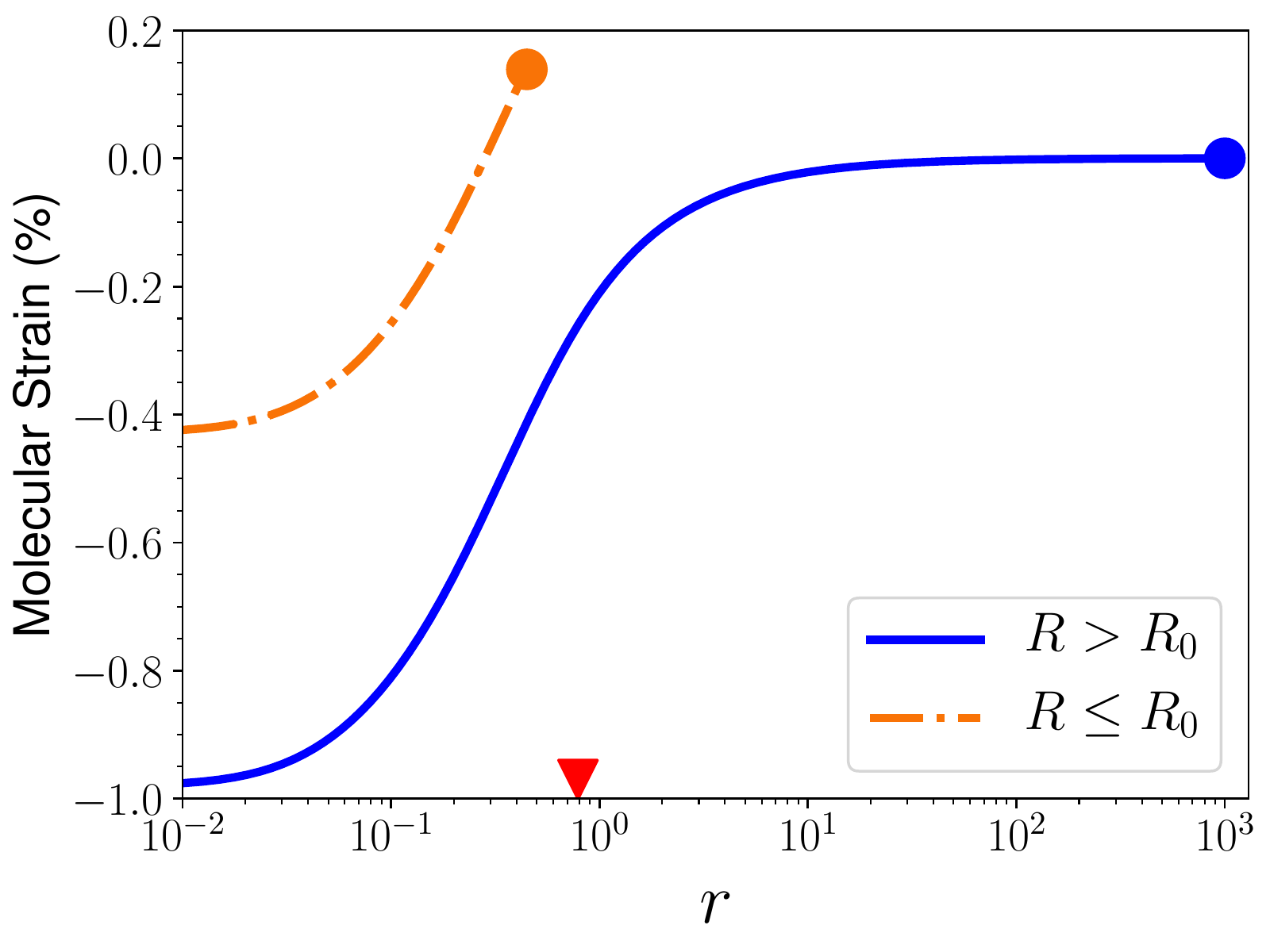}
  \caption{Molecular strain $\epsilon(r)$ throughout the fibril. We show fibrils with both $R \lesssim R_0$ (orange dash-dotted) and $R > R_0$ (blue solid). Coloured circles represent the fibril radius, while the red triangle indicates $R_0$. Parameters used are the same as in Fig.~\ref{fig:solution}.}
  \label{fig:molecularstrain}
\end{figure}

In order to maintain a global D-band throughout the fibril, we have assumed that collagen molecules are subject to small-scale extension and/or compression along their axis within the fibril. Fig.~\ref{fig:molecularstrain} shows the resulting local molecular strain $\%$, defined as \cite{Cameron:2020}
\begin{equation}
    \epsilon(r) = 100\times \frac{2\pi/\eta(R)-\cos\psi(r)}{\cos\psi(r)},
\end{equation}
where $\eta(R)$ is the global D-band wavenumber for a fibril of radius $R$ while $\psi(r)$ is the double-twist function of the molecular tilt within the fibril. The absolute molecular strain is at most a couple of percent which, given the flexibility observed in experimental studies of isolated collagen fibrils,\cite{Rezaei:2018} appears reasonable.

For fibrils in the $R > R_0$ regime, the interior of the fibril is entirely compressed -- with a negative strain (blue line in Fig.~\ref{fig:molecularstrain}). We find that the molecules in the core of large fibrils are compressed by up to $1\%$. Compression of the entire fibril is a qualitatively new result from the nonequilibrium growth model. In contrast, earlier equilibrium models of \textit{in vitro} D-banded always predicted an extended surface layer in addition to the compressed core.\cite{Cameron:2020} For our non-equilibrium model an extended surface is only seen for fibrils with $R \lesssim R_0$ (dot-dashed orange line in Fig.~\ref{fig:molecularstrain}). The molecular compression shown in Fig.~\ref{fig:molecularstrain} is similar to the hypothesis of molecular microkinks proposed to explain diffuse x-ray scattering in tendon studies,\cite{Misof:1997} though our strains are somewhat smaller than they suggest. We speculate that pre-compression may serve to protect tendon-like fibrils from damage at small but physiologically relevant, extensional strains.

\section{Appendix: Stable Radius Control}
\label{sec:RadiusControl}
Experiment suggests that the fractional variation  of fibril radius near the center of the cornea is only $\sigma_{\tilde{R}}/\tilde{R} \simeq 0.16$.\cite{Cox:1970} Here we estimate the maximum concentration variation needed to achieve this control of corneal radius, using the stable-fixed point illustrated in Fig.~\ref{fig:radius_control} but not requiring that $R \approx R_0$.  From equation~\ref{stable} we have
\begin{eqnarray}
    \frac{\Delta n}{n} &\approx& \tilde{f}' \tilde{R}
             \frac{\sigma_{\tilde{R}}/ \tilde{R}}{n_f k_B T} \\
    & \approx& f' R
             \frac{\sigma_{R}/R}{n_f k_B T} \frac{\tilde{K}_{22} q^2 (1+k_{24})}{1-k_{24}},
             \label{dnn}
\end{eqnarray}
where we have used a dimensionless $f'$ in the second line.  We use $k_B T=4.28 \times 10^{-21}$J.

\begin{figure}[t!] 
   \centering
   \includegraphics[width=0.4\textwidth]{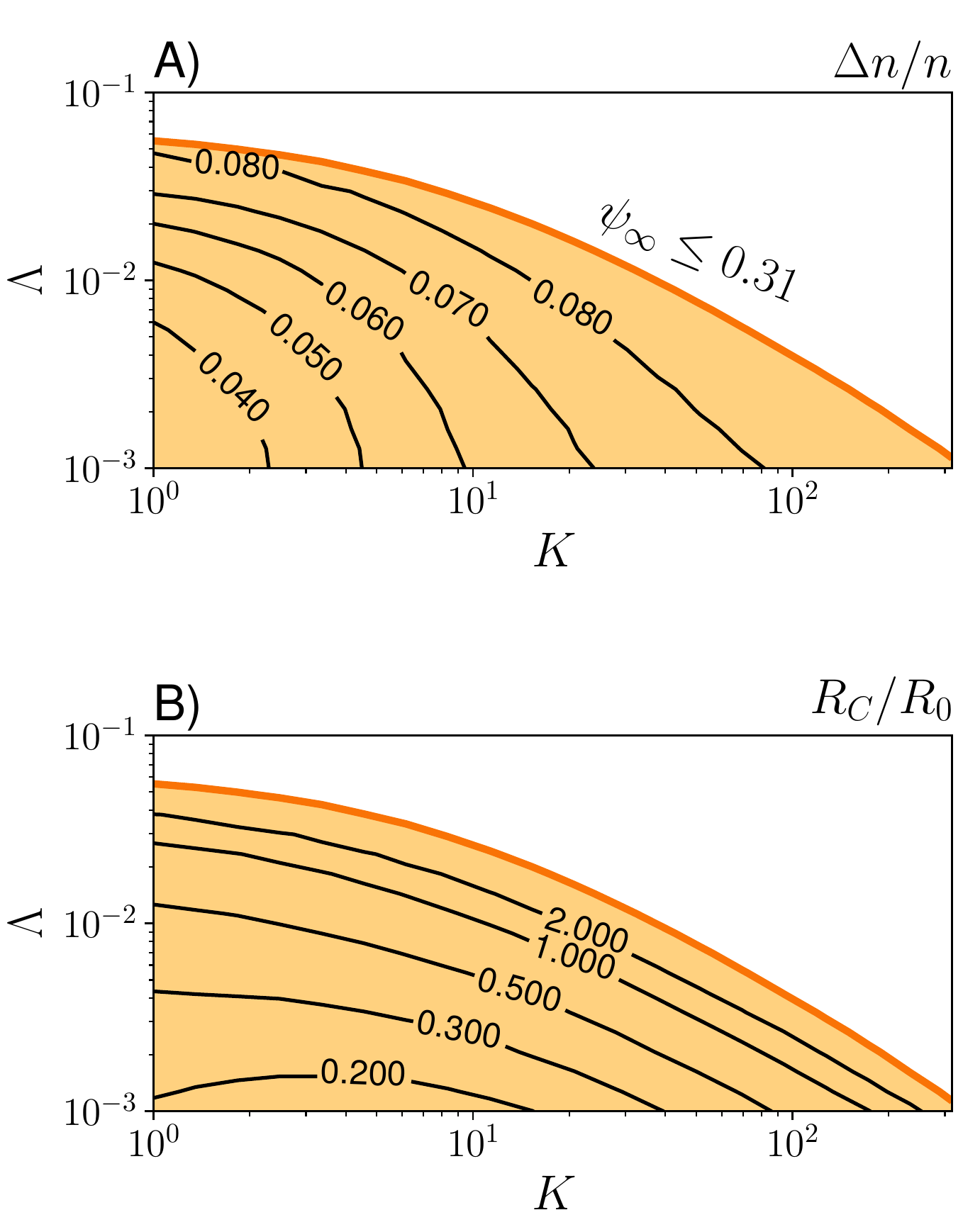} 
   \caption{A) $\Delta n/n$ relative concentration variations consistent with observed corneal radius variations (see Eqn.~\ref{dnn}) at $R_C$ and $\psi_C(R_C)$ vs $K$ and $\Lambda$.
   B) Corresponding $R_C/R_0$. For A) we use $\tilde{K}_{22}=6$pN, $q=5\pi\mu$m$^{-1}$, and $k_{24}^C$ as in Fig.~\ref{fig:cornea}A. }
   \label{fig:deltann}
\end{figure}

In Fig.~\ref{fig:deltann}A we plot $\Delta n/n$ as a function of $K$ and $\Lambda$. We have  used $k_{24}^C$ from figure~\ref{fig:cornea}C, $q=5\pi \mu$m$^{-1}$, and $\tilde{K}_{22} \approx 6$pN at the upper end of its estimated range.\cite{Cameron:2018} We see that the largest concentration variations  consistent for the observed corneal radius control are $\Delta n/n \lesssim 10\%$. This seems plausible. In  Fig.~\ref{fig:deltann}B we show that $R_C/R_0 \approx 1$ leads to the loosest limits on concentration variations. 

\section*{Conflicts of interest}
There are no conflicts to declare.

\section*{Acknowledgements}
We thank the Natural Sciences and Engineering Research Council of Canada (NSERC) for operating Grants RGPIN-2018-03781 (LK) and RGPIN-2019-05888 (ADR). MPL thanks NSERC for summer fellowship support (USRA-552365-2020).

\renewcommand\refname{Notes and references} 
\bibliography{main} 

\providecommand*{\mcitethebibliography}{\thebibliography}
\csname @ifundefined\endcsname{endmcitethebibliography}
{\let\endmcitethebibliography\endthebibliography}{}
\begin{mcitethebibliography}{70}
\providecommand*{\natexlab}[1]{#1}
\providecommand*{\mciteSetBstSublistMode}[1]{}
\providecommand*{\mciteSetBstMaxWidthForm}[2]{}
\providecommand*{\mciteBstWouldAddEndPuncttrue}
  {\def\EndOfBibitem{\unskip.}}
\providecommand*{\mciteBstWouldAddEndPunctfalse}
  {\let\EndOfBibitem\relax}
\providecommand*{\mciteSetBstMidEndSepPunct}[3]{}
\providecommand*{\mciteSetBstSublistLabelBeginEnd}[3]{}
\providecommand*{\EndOfBibitem}{}
\mciteSetBstSublistMode{f}
\mciteSetBstMaxWidthForm{subitem}
{(\emph{\alph{mcitesubitemcount}})}
\mciteSetBstSublistLabelBeginEnd{\mcitemaxwidthsubitemform\space}
{\relax}{\relax}

\bibitem[Ricard-Blum(2011)]{Ricard-Blum:2011}
S.~Ricard-Blum, \emph{Cold Spring Harbor Perspectives in Biology}, 2011,
  \textbf{3}, a004978\relax
\mciteBstWouldAddEndPuncttrue
\mciteSetBstMidEndSepPunct{\mcitedefaultmidpunct}
{\mcitedefaultendpunct}{\mcitedefaultseppunct}\relax
\EndOfBibitem
\bibitem[Sherman \emph{et~al.}(2015)Sherman, Yang, and Meyers]{Sherman:2015}
V.~R. Sherman, W.~Yang and M.~A. Meyers, \emph{Journal of the Mechanical
  Behavior of Biomedical Materials}, 2015, \textbf{52}, 22--50\relax
\mciteBstWouldAddEndPuncttrue
\mciteSetBstMidEndSepPunct{\mcitedefaultmidpunct}
{\mcitedefaultendpunct}{\mcitedefaultseppunct}\relax
\EndOfBibitem
\bibitem[Rezaei \emph{et~al.}(2018)Rezaei, Lyons, and Forde]{Rezaei:2018}
N.~Rezaei, A.~Lyons and N.~R. Forde, \emph{Biophysical Journal}, 2018,
  \textbf{115}, 1457--1469\relax
\mciteBstWouldAddEndPuncttrue
\mciteSetBstMidEndSepPunct{\mcitedefaultmidpunct}
{\mcitedefaultendpunct}{\mcitedefaultseppunct}\relax
\EndOfBibitem
\bibitem[Fang \emph{et~al.}(2012)Fang, Goldstein, Turner, Les, Orr, Fisher,
  Welch, Rothman, and Banaszak~Holl]{Fang:2012}
M.~Fang, E.~L. Goldstein, A.~S. Turner, C.~M. Les, B.~G. Orr, G.~J. Fisher,
  K.~B. Welch, E.~D. Rothman and M.~M. Banaszak~Holl, \emph{ACS Nano}, 2012,
  \textbf{6}, 9503--9514\relax
\mciteBstWouldAddEndPuncttrue
\mciteSetBstMidEndSepPunct{\mcitedefaultmidpunct}
{\mcitedefaultendpunct}{\mcitedefaultseppunct}\relax
\EndOfBibitem
\bibitem[Quan and Sone(2015)]{Quan:2015}
B.~D. Quan and E.~D. Sone, \emph{Bone}, 2015, \textbf{77}, 42--49\relax
\mciteBstWouldAddEndPuncttrue
\mciteSetBstMidEndSepPunct{\mcitedefaultmidpunct}
{\mcitedefaultendpunct}{\mcitedefaultseppunct}\relax
\EndOfBibitem
\bibitem[Raspanti \emph{et~al.}(2018)Raspanti, Reguzzoni, Protasoni, and
  Basso]{Raspanti:2018}
M.~Raspanti, M.~Reguzzoni, M.~Protasoni and P.~Basso, \emph{International
  Journal of Biological Macromolecules}, 2018, \textbf{107}, 1668--1674\relax
\mciteBstWouldAddEndPuncttrue
\mciteSetBstMidEndSepPunct{\mcitedefaultmidpunct}
{\mcitedefaultendpunct}{\mcitedefaultseppunct}\relax
\EndOfBibitem
\bibitem[Holmes \emph{et~al.}(2001)Holmes, Gilpin, Baldock, Ziese, Koster, and
  Kadler]{Holmes:2001}
D.~F. Holmes, C.~J. Gilpin, C.~Baldock, U.~Ziese, A.~J. Koster and K.~E.
  Kadler, \emph{Proceedings of the National Academy of Sciences}, 2001,
  \textbf{98}, 7307--7312\relax
\mciteBstWouldAddEndPuncttrue
\mciteSetBstMidEndSepPunct{\mcitedefaultmidpunct}
{\mcitedefaultendpunct}{\mcitedefaultseppunct}\relax
\EndOfBibitem
\bibitem[Bell \emph{et~al.}(2018)Bell, Hayes, Whitford, Sanchez-Weatherby,
  Shebanova, Vergari, Winlove, Terrill, Sorensen, Elsheikh, and
  Meek]{Bell:2018}
J.~S. Bell, S.~Hayes, C.~Whitford, J.~Sanchez-Weatherby, O.~Shebanova,
  C.~Vergari, C.~P. Winlove, N.~Terrill, T.~Sorensen, A.~Elsheikh and K.~M.
  Meek, \emph{Acta Biomaterialia}, 2018, \textbf{65}, 216--225\relax
\mciteBstWouldAddEndPuncttrue
\mciteSetBstMidEndSepPunct{\mcitedefaultmidpunct}
{\mcitedefaultendpunct}{\mcitedefaultseppunct}\relax
\EndOfBibitem
\bibitem[Meek and Knupp(2015)]{Meek:2015}
K.~M. Meek and C.~Knupp, \emph{Progress in Retinal and Eye Research}, 2015,
  \textbf{49}, 1--16\relax
\mciteBstWouldAddEndPuncttrue
\mciteSetBstMidEndSepPunct{\mcitedefaultmidpunct}
{\mcitedefaultendpunct}{\mcitedefaultseppunct}\relax
\EndOfBibitem
\bibitem[Parry \emph{et~al.}(1978)Parry, Barnes, and Craig]{Parry:1978}
D.~A. Parry, G.~R. Barnes and A.~S. Craig, \emph{Proc. R. Soc. London B}, 1978,
  \textbf{203}, 305--321\relax
\mciteBstWouldAddEndPuncttrue
\mciteSetBstMidEndSepPunct{\mcitedefaultmidpunct}
{\mcitedefaultendpunct}{\mcitedefaultseppunct}\relax
\EndOfBibitem
\bibitem[Patterson-Kane \emph{et~al.}(1997)Patterson-Kane, Wilson, Firth,
  Parry, and Goodship]{PattersonKane:1997}
J.~C. Patterson-Kane, A.~M. Wilson, E.~C. Firth, D.~A. Parry and A.~E.
  Goodship, \emph{Equine Veterinary Journal}, 1997, \textbf{29}, 121--125\relax
\mciteBstWouldAddEndPuncttrue
\mciteSetBstMidEndSepPunct{\mcitedefaultmidpunct}
{\mcitedefaultendpunct}{\mcitedefaultseppunct}\relax
\EndOfBibitem
\bibitem[Goh \emph{et~al.}(2012)Goh, Holmes, Lu, Purslow, Kadler, Bechet, and
  Wess]{Goh:2012}
K.~L. Goh, D.~F. Holmes, Y.~Lu, P.~P. Purslow, K.~E. Kadler, D.~Bechet and
  T.~J. Wess, \emph{Journal of Applied Physiology}, 2012, \textbf{113},
  878--888\relax
\mciteBstWouldAddEndPuncttrue
\mciteSetBstMidEndSepPunct{\mcitedefaultmidpunct}
{\mcitedefaultendpunct}{\mcitedefaultseppunct}\relax
\EndOfBibitem
\bibitem[Kalson \emph{et~al.}(2015)Kalson, Lu, Taylor, Starborg, Holmes, and
  Kadler]{Kalson:2015}
N.~S. Kalson, Y.~Lu, S.~H. Taylor, T.~Starborg, D.~F. Holmes and K.~E. Kadler,
  \emph{eLife}, 2015, \textbf{4}, e05958\relax
\mciteBstWouldAddEndPuncttrue
\mciteSetBstMidEndSepPunct{\mcitedefaultmidpunct}
{\mcitedefaultendpunct}{\mcitedefaultseppunct}\relax
\EndOfBibitem
\bibitem[Reale \emph{et~al.}(1981)Reale, Benazzo, and Ruggeri]{Reale:1981}
E.~Reale, F.~Benazzo and A.~Ruggeri, \emph{Journal of Submicroscopic Cytology},
  1981, \textbf{13}, 135--143\relax
\mciteBstWouldAddEndPuncttrue
\mciteSetBstMidEndSepPunct{\mcitedefaultmidpunct}
{\mcitedefaultendpunct}{\mcitedefaultseppunct}\relax
\EndOfBibitem
\bibitem[Zhou \emph{et~al.}(2016)Zhou, Burger, Wang, Hsiao, Chu, and
  Graham]{Zhou:2016}
H.-W. Zhou, C.~Burger, H.~Wang, B.~S. Hsiao, B.~Chu and L.~Graham, \emph{Acta
  Cryst. D}, 2016, \textbf{72}, 986--996\relax
\mciteBstWouldAddEndPuncttrue
\mciteSetBstMidEndSepPunct{\mcitedefaultmidpunct}
{\mcitedefaultendpunct}{\mcitedefaultseppunct}\relax
\EndOfBibitem
\bibitem[Xu \emph{et~al.}(2018)Xu, Zhao, Wang, Yang, and Sahai]{Xu:2018}
Z.~Xu, W.~Zhao, Z.~Wang, Y.~Yang and N.~Sahai, \emph{Physical Chemistry
  Chemical Physics}, 2018, \textbf{20}, 1513--1523\relax
\mciteBstWouldAddEndPuncttrue
\mciteSetBstMidEndSepPunct{\mcitedefaultmidpunct}
{\mcitedefaultendpunct}{\mcitedefaultseppunct}\relax
\EndOfBibitem
\bibitem[Brown \emph{et~al.}(2014)Brown, Kreplak, and Rutenberg]{Brown:2014}
A.~I. Brown, L.~Kreplak and A.~D. Rutenberg, \emph{Soft Matter}, 2014,
  \textbf{10}, 8500--8511\relax
\mciteBstWouldAddEndPuncttrue
\mciteSetBstMidEndSepPunct{\mcitedefaultmidpunct}
{\mcitedefaultendpunct}{\mcitedefaultseppunct}\relax
\EndOfBibitem
\bibitem[Cameron \emph{et~al.}(2018)Cameron, Kreplak, and
  Rutenberg]{Cameron:2018}
S.~Cameron, L.~Kreplak and A.~D. Rutenberg, \emph{Soft Matter}, 2018,
  \textbf{14}, 4772--4783\relax
\mciteBstWouldAddEndPuncttrue
\mciteSetBstMidEndSepPunct{\mcitedefaultmidpunct}
{\mcitedefaultendpunct}{\mcitedefaultseppunct}\relax
\EndOfBibitem
\bibitem[Cameron \emph{et~al.}(2020)Cameron, Kreplak, and
  Rutenberg]{Cameron:2020}
S.~Cameron, L.~Kreplak and A.~D. Rutenberg, \emph{Physical Review Research},
  2020, \textbf{2}, 012070(R)\relax
\mciteBstWouldAddEndPuncttrue
\mciteSetBstMidEndSepPunct{\mcitedefaultmidpunct}
{\mcitedefaultendpunct}{\mcitedefaultseppunct}\relax
\EndOfBibitem
\bibitem[Harris \emph{et~al.}(2013)Harris, Soliakov, and Lewis]{Harris:2013}
J.~R. Harris, A.~Soliakov and R.~J. Lewis, \emph{Micron}, 2013, \textbf{49},
  60--68\relax
\mciteBstWouldAddEndPuncttrue
\mciteSetBstMidEndSepPunct{\mcitedefaultmidpunct}
{\mcitedefaultendpunct}{\mcitedefaultseppunct}\relax
\EndOfBibitem
\bibitem[Gobeaux \emph{et~al.}(2008)Gobeaux, Mosser, Anglo, Panine, Davidson,
  Giraud-Guille, and Belamie]{Gobeaux:2008}
F.~Gobeaux, G.~Mosser, A.~Anglo, P.~Panine, P.~Davidson, M.~M. Giraud-Guille
  and E.~Belamie, \emph{Journal of Molecular Biology}, 2008, \textbf{376},
  1509--1522\relax
\mciteBstWouldAddEndPuncttrue
\mciteSetBstMidEndSepPunct{\mcitedefaultmidpunct}
{\mcitedefaultendpunct}{\mcitedefaultseppunct}\relax
\EndOfBibitem
\bibitem[Kagan \emph{et~al.}(1991)Kagan, Trackman,\emph{et~al.}]{Kagan:1991}
H.~M. Kagan, P.~C. Trackman \emph{et~al.}, \emph{Am J Respir Cell Mol Biol},
  1991, \textbf{5}, 206--210\relax
\mciteBstWouldAddEndPuncttrue
\mciteSetBstMidEndSepPunct{\mcitedefaultmidpunct}
{\mcitedefaultendpunct}{\mcitedefaultseppunct}\relax
\EndOfBibitem
\bibitem[Eekhoff \emph{et~al.}(2018)Eekhoff, Fang, and Lake]{Eekhoff:2018}
J.~D. Eekhoff, F.~Fang and S.~P. Lake, \emph{Connective Tissue Research}, 2018,
  \textbf{59}, 410--422\relax
\mciteBstWouldAddEndPuncttrue
\mciteSetBstMidEndSepPunct{\mcitedefaultmidpunct}
{\mcitedefaultendpunct}{\mcitedefaultseppunct}\relax
\EndOfBibitem
\bibitem[Makris \emph{et~al.}(2014)Makris, Responte, Paschos, Hu, and
  Athanasiou]{Makris:2014}
E.~A. Makris, D.~J. Responte, N.~K. Paschos, J.~C. Hu and K.~A. Athanasiou,
  \emph{Proceedings of the National Academy of Sciences}, 2014, \textbf{111},
  E4832--E4841\relax
\mciteBstWouldAddEndPuncttrue
\mciteSetBstMidEndSepPunct{\mcitedefaultmidpunct}
{\mcitedefaultendpunct}{\mcitedefaultseppunct}\relax
\EndOfBibitem
\bibitem[Depalle \emph{et~al.}(2015)Depalle, Qin, Shefelbine, and
  Buehler]{Depalle:2015}
B.~Depalle, Z.~Qin, S.~J. Shefelbine and M.~J. Buehler, \emph{Journal of the
  Mechanical Behavior of Biomedical Materials}, 2015, \textbf{52}, 1--13\relax
\mciteBstWouldAddEndPuncttrue
\mciteSetBstMidEndSepPunct{\mcitedefaultmidpunct}
{\mcitedefaultendpunct}{\mcitedefaultseppunct}\relax
\EndOfBibitem
\bibitem[Ekani-Nkodo and Fygenson(2003)]{EkaniNkodo:2003}
A.~Ekani-Nkodo and D.~K. Fygenson, \emph{Physical Review E}, 2003, \textbf{67},
  021909--7\relax
\mciteBstWouldAddEndPuncttrue
\mciteSetBstMidEndSepPunct{\mcitedefaultmidpunct}
{\mcitedefaultendpunct}{\mcitedefaultseppunct}\relax
\EndOfBibitem
\bibitem[Toroian \emph{et~al.}(2007)Toroian, Lim, and Price]{Toroian:2007}
D.~Toroian, J.~E. Lim and P.~A. Price, \emph{Journal of Biological Chemistry},
  2007, \textbf{282}, 22437--22447\relax
\mciteBstWouldAddEndPuncttrue
\mciteSetBstMidEndSepPunct{\mcitedefaultmidpunct}
{\mcitedefaultendpunct}{\mcitedefaultseppunct}\relax
\EndOfBibitem
\bibitem[Vallet \emph{et~al.}(2018)Vallet, Miele, Uciechowska-Kaczmarzyk, Liwo,
  Duclos, Samsonov, and Ricard-Blum]{Vallet:2018}
S.~D. Vallet, A.~E. Miele, U.~Uciechowska-Kaczmarzyk, A.~Liwo, B.~Duclos, S.~A.
  Samsonov and S.~Ricard-Blum, \emph{Scientific Reports}, 2018, \textbf{8},
  1--16\relax
\mciteBstWouldAddEndPuncttrue
\mciteSetBstMidEndSepPunct{\mcitedefaultmidpunct}
{\mcitedefaultendpunct}{\mcitedefaultseppunct}\relax
\EndOfBibitem
\bibitem[Xu \emph{et~al.}(2020)Xu, Nudelman, Eren, Wirix, Cantaert, Nijhuis,
  Hermida-Merino, Portale, Bomans, Ottmann, Friedrich, Bras, Akiva, Orgel,
  Meldrum, and Sommerdijk]{Xu:2020}
Y.~Xu, F.~Nudelman, E.~D. Eren, M.~J.~M. Wirix, B.~Cantaert, W.~H. Nijhuis,
  D.~Hermida-Merino, G.~Portale, P.~H.~H. Bomans, C.~Ottmann, H.~Friedrich,
  W.~Bras, A.~Akiva, J.~P. R.~O. Orgel, F.~C. Meldrum and N.~Sommerdijk,
  \emph{Nature Communications}, 2020, \textbf{11}, 5068\relax
\mciteBstWouldAddEndPuncttrue
\mciteSetBstMidEndSepPunct{\mcitedefaultmidpunct}
{\mcitedefaultendpunct}{\mcitedefaultseppunct}\relax
\EndOfBibitem
\bibitem[Marturano \emph{et~al.}(2014)Marturano, Xylas, Sridharan, Georgakoudi,
  and Kuo]{Marturano:2014}
J.~E. Marturano, J.~F. Xylas, G.~V. Sridharan, I.~Georgakoudi and C.~K. Kuo,
  \emph{Acta Biomaterialia}, 2014, \textbf{10}, 1370--1379\relax
\mciteBstWouldAddEndPuncttrue
\mciteSetBstMidEndSepPunct{\mcitedefaultmidpunct}
{\mcitedefaultendpunct}{\mcitedefaultseppunct}\relax
\EndOfBibitem
\bibitem[Herchenhan \emph{et~al.}(2015)Herchenhan, Uhlenbrock, Eliasson, Weis,
  Eyre, Kadler, Magnusson, and Kjaer]{Herchenhan:2015}
A.~Herchenhan, F.~Uhlenbrock, P.~Eliasson, M.~Weis, D.~Eyre, K.~E. Kadler,
  S.~P. Magnusson and M.~Kjaer, \emph{Journal of Biological Chemistry}, 2015,
  \textbf{290}, 16440--16450\relax
\mciteBstWouldAddEndPuncttrue
\mciteSetBstMidEndSepPunct{\mcitedefaultmidpunct}
{\mcitedefaultendpunct}{\mcitedefaultseppunct}\relax
\EndOfBibitem
\bibitem[Hulmes \emph{et~al.}(1995)Hulmes, Wess, Prockop, and
  Fratzl]{Hulmes:1995}
D.~J. Hulmes, T.~J. Wess, D.~J. Prockop and P.~Fratzl, \emph{Biophysical
  Journal}, 1995, \textbf{68}, 1661--1670\relax
\mciteBstWouldAddEndPuncttrue
\mciteSetBstMidEndSepPunct{\mcitedefaultmidpunct}
{\mcitedefaultendpunct}{\mcitedefaultseppunct}\relax
\EndOfBibitem
\bibitem[Baldwin \emph{et~al.}(2020)Baldwin, Sampson, Peacock, Martin, Veres,
  Lee, and Kreplak]{Baldwin:2020}
S.~J. Baldwin, J.~Sampson, C.~J. Peacock, M.~L. Martin, S.~P. Veres, J.~M. Lee
  and L.~Kreplak, \emph{Journal of the Mechanical Behavior of Biomedical
  Materials}, 2020, \textbf{110}, 103849\relax
\mciteBstWouldAddEndPuncttrue
\mciteSetBstMidEndSepPunct{\mcitedefaultmidpunct}
{\mcitedefaultendpunct}{\mcitedefaultseppunct}\relax
\EndOfBibitem
\bibitem[Berenguer \emph{et~al.}(2014)Berenguer, Bean, Bozec, Vila-Comamala,
  Zhang, Kewish, Bunk, Rodenburg, and Robinson]{Berenguer:2014}
F.~Berenguer, R.~J. Bean, L.~Bozec, J.~Vila-Comamala, F.~Zhang, C.~M. Kewish,
  O.~Bunk, J.~M. Rodenburg and I.~K. Robinson, \emph{Biophysical Journal},
  2014, \textbf{106}, 459--466\relax
\mciteBstWouldAddEndPuncttrue
\mciteSetBstMidEndSepPunct{\mcitedefaultmidpunct}
{\mcitedefaultendpunct}{\mcitedefaultseppunct}\relax
\EndOfBibitem
\bibitem[Hodge(1989)]{Hodge:1989}
A.~J. Hodge, \emph{Connective Tissue Research}, 1989, \textbf{21},
  137--147\relax
\mciteBstWouldAddEndPuncttrue
\mciteSetBstMidEndSepPunct{\mcitedefaultmidpunct}
{\mcitedefaultendpunct}{\mcitedefaultseppunct}\relax
\EndOfBibitem
\bibitem[Ericksen(1991)]{Ericksen:1991}
J.~Ericksen, \emph{Archive for Rational Mechanics and Analysis}, 1991,
  \textbf{113}, 97--120\relax
\mciteBstWouldAddEndPuncttrue
\mciteSetBstMidEndSepPunct{\mcitedefaultmidpunct}
{\mcitedefaultendpunct}{\mcitedefaultseppunct}\relax
\EndOfBibitem
\bibitem[git()]{github}
\emph{Github code},
  \url{https://github.com/Matthew-Leighton/Nonequilibrium_Fibril_Growth}\relax
\mciteBstWouldAddEndPuncttrue
\mciteSetBstMidEndSepPunct{\mcitedefaultmidpunct}
{\mcitedefaultendpunct}{\mcitedefaultseppunct}\relax
\EndOfBibitem
\bibitem[Gray(1959)]{Gray:1959}
E.~G. Gray, \emph{Proceedings of the Royal Society of London. Series B}, 1959,
  \textbf{150}, 233--239\relax
\mciteBstWouldAddEndPuncttrue
\mciteSetBstMidEndSepPunct{\mcitedefaultmidpunct}
{\mcitedefaultendpunct}{\mcitedefaultseppunct}\relax
\EndOfBibitem
\bibitem[Boatman \emph{et~al.}(2019)Boatman, Goodwin, Holman, Fakra, Zheng,
  Gronsky, and Schweitzer]{Boatman:2019}
E.~M. Boatman, M.~B. Goodwin, H.-Y.~N. Holman, S.~Fakra, W.~Zheng, R.~Gronsky
  and M.~H. Schweitzer, \emph{Scientific Reports}, 2019, \textbf{9},
  1--12\relax
\mciteBstWouldAddEndPuncttrue
\mciteSetBstMidEndSepPunct{\mcitedefaultmidpunct}
{\mcitedefaultendpunct}{\mcitedefaultseppunct}\relax
\EndOfBibitem
\bibitem[Yamamoto \emph{et~al.}(2000)Yamamoto, Hashizume, Hitomi, Shigeno,
  Sawaguchi, Abe, and Ushiki]{Yamamoto:2000}
S.~Yamamoto, H.~Hashizume, J.~Hitomi, M.~Shigeno, S.~Sawaguchi, H.~Abe and
  T.~Ushiki, \emph{Archives of Histology}, 2000, \textbf{63}, 127--135\relax
\mciteBstWouldAddEndPuncttrue
\mciteSetBstMidEndSepPunct{\mcitedefaultmidpunct}
{\mcitedefaultendpunct}{\mcitedefaultseppunct}\relax
\EndOfBibitem
\bibitem[Hirsch \emph{et~al.}(2001)Hirsch, Prenant, and Renard]{Hirsch:2001}
M.~Hirsch, G.~Prenant and G.~Renard, \emph{Experimental Eye Research}, 2001,
  \textbf{72}, 123--135\relax
\mciteBstWouldAddEndPuncttrue
\mciteSetBstMidEndSepPunct{\mcitedefaultmidpunct}
{\mcitedefaultendpunct}{\mcitedefaultseppunct}\relax
\EndOfBibitem
\bibitem[Yamamoto \emph{et~al.}(1997)Yamamoto, Hitomi, Shigeno, Sawaguchi, Abe,
  and Ushiki]{Yamamoto:1997}
S.~Yamamoto, J.~Hitomi, M.~Shigeno, S.~Sawaguchi, H.~Abe and T.~Ushiki,
  \emph{Archives of Histology and Cytology}, 1997, \textbf{60}, 371--378\relax
\mciteBstWouldAddEndPuncttrue
\mciteSetBstMidEndSepPunct{\mcitedefaultmidpunct}
{\mcitedefaultendpunct}{\mcitedefaultseppunct}\relax
\EndOfBibitem
\bibitem[Quigley \emph{et~al.}(2018)Quigley, Bancelin, Deska-Gauthier,
  L{\'e}gar{\'e}, Kreplak, and Veres]{Quigley:2018}
A.~S. Quigley, S.~Bancelin, D.~Deska-Gauthier, F.~L{\'e}gar{\'e}, L.~Kreplak
  and S.~P. Veres, \emph{Scientific Reports}, 2018, \textbf{8}, 4409\relax
\mciteBstWouldAddEndPuncttrue
\mciteSetBstMidEndSepPunct{\mcitedefaultmidpunct}
{\mcitedefaultendpunct}{\mcitedefaultseppunct}\relax
\EndOfBibitem
\bibitem[Baselt \emph{et~al.}(1993)Baselt, Revel, and
  Baldeschwieler]{Baselt:1993}
D.~R. Baselt, J.~P. Revel and J.~D. Baldeschwieler, \emph{Biophysical Journal},
  1993, \textbf{65}, 2644--2655\relax
\mciteBstWouldAddEndPuncttrue
\mciteSetBstMidEndSepPunct{\mcitedefaultmidpunct}
{\mcitedefaultendpunct}{\mcitedefaultseppunct}\relax
\EndOfBibitem
\bibitem[rat()]{ratchet}
Any slow irreversible attachment against the chemical potential gradient due to
  cross-linking is ignored here.\relax
\mciteBstWouldAddEndPunctfalse
\mciteSetBstMidEndSepPunct{\mcitedefaultmidpunct}
{}{\mcitedefaultseppunct}\relax
\EndOfBibitem
\bibitem[Rutenberg \emph{et~al.}(2016)Rutenberg, Brown, and
  Kreplak]{Rutenberg:2016}
A.~D. Rutenberg, A.~I. Brown and L.~Kreplak, \emph{Physical Biology}, 2016,
  \textbf{13}, 046008\relax
\mciteBstWouldAddEndPuncttrue
\mciteSetBstMidEndSepPunct{\mcitedefaultmidpunct}
{\mcitedefaultendpunct}{\mcitedefaultseppunct}\relax
\EndOfBibitem
\bibitem[De~Sa~Peixoto \emph{et~al.}(2011)De~Sa~Peixoto, Deniset-Besseau,
  Schanne-Klein, and Mosser]{DeSaPeixoto:2011}
P.~De~Sa~Peixoto, A.~Deniset-Besseau, M.-C. Schanne-Klein and G.~Mosser,
  \emph{Soft Matter}, 2011, \textbf{7}, 11203--11210\relax
\mciteBstWouldAddEndPuncttrue
\mciteSetBstMidEndSepPunct{\mcitedefaultmidpunct}
{\mcitedefaultendpunct}{\mcitedefaultseppunct}\relax
\EndOfBibitem
\bibitem[Cox \emph{et~al.}(1970)Cox, Farrell, Hart, and Langham]{Cox:1970}
J.~L. Cox, R.~A. Farrell, R.~W. Hart and M.~E. Langham, \emph{Journal of
  Physiology}, 1970, \textbf{210}, 601--616\relax
\mciteBstWouldAddEndPuncttrue
\mciteSetBstMidEndSepPunct{\mcitedefaultmidpunct}
{\mcitedefaultendpunct}{\mcitedefaultseppunct}\relax
\EndOfBibitem
\bibitem[Warner and Terentjev(1996)]{Warner:1996}
M.~Warner and E.~Terentjev, \emph{Progress in Polymer Science}, 1996,
  \textbf{21}, 853--891\relax
\mciteBstWouldAddEndPuncttrue
\mciteSetBstMidEndSepPunct{\mcitedefaultmidpunct}
{\mcitedefaultendpunct}{\mcitedefaultseppunct}\relax
\EndOfBibitem
\bibitem[Saito \emph{et~al.}(1997)Saito, Marumo, Fujii, and
  Ishioka]{Saito:1997}
M.~Saito, K.~Marumo, K.~Fujii and N.~Ishioka, \emph{Analytical Biochemistry},
  1997, \textbf{253}, 26--32\relax
\mciteBstWouldAddEndPuncttrue
\mciteSetBstMidEndSepPunct{\mcitedefaultmidpunct}
{\mcitedefaultendpunct}{\mcitedefaultseppunct}\relax
\EndOfBibitem
\bibitem[equ()]{equilibrium}
A fully equilibrium model of fibril structure can exhibit a broad coexistence
  regime between small and large radius fibrils.\cite{Cameron:2020} In our
  non-equilibrium model a broad range of radii is a non-equilibrium effect, due
  to crude control of fibril growth for $R \gg R_0$.\relax
\mciteBstWouldAddEndPunctfalse
\mciteSetBstMidEndSepPunct{\mcitedefaultmidpunct}
{}{\mcitedefaultseppunct}\relax
\EndOfBibitem
\bibitem[Gutsmann \emph{et~al.}(2003)Gutsmann, Fantner, Venturoni, journal, and
  {2003}]{Gutsmann:2003}
T.~Gutsmann, G.~E. Fantner, M.~Venturoni, A.~E.-N.~B. journal and {2003},
  \emph{Biophysical Journal}, 2003, \textbf{84}, 2593--2598\relax
\mciteBstWouldAddEndPuncttrue
\mciteSetBstMidEndSepPunct{\mcitedefaultmidpunct}
{\mcitedefaultendpunct}{\mcitedefaultseppunct}\relax
\EndOfBibitem
\bibitem[Wenger \emph{et~al.}(2008)Wenger, Horton, and Mesquida]{Wenger:2008}
M.~P.~E. Wenger, M.~A. Horton and P.~Mesquida, \emph{Nanotechnology}, 2008,
  \textbf{19}, 384006\relax
\mciteBstWouldAddEndPuncttrue
\mciteSetBstMidEndSepPunct{\mcitedefaultmidpunct}
{\mcitedefaultendpunct}{\mcitedefaultseppunct}\relax
\EndOfBibitem
\bibitem[Strasser \emph{et~al.}(2007)Strasser, Zink, Janko, Heckl, and
  Thalhammer]{Strasser:2007}
S.~Strasser, A.~Zink, M.~Janko, W.~M. Heckl and S.~Thalhammer,
  \emph{Biochemical and Biophysical Research Communications}, 2007,
  \textbf{354}, 27--32\relax
\mciteBstWouldAddEndPuncttrue
\mciteSetBstMidEndSepPunct{\mcitedefaultmidpunct}
{\mcitedefaultendpunct}{\mcitedefaultseppunct}\relax
\EndOfBibitem
\bibitem[Slatter \emph{et~al.}(2008)Slatter, Avery, and Bailey]{Slatter:2008}
D.~A. Slatter, N.~C. Avery and A.~J. Bailey, \emph{International Journal of
  Biochemistry and Cell Biology}, 2008, \textbf{40}, 2253--2263\relax
\mciteBstWouldAddEndPuncttrue
\mciteSetBstMidEndSepPunct{\mcitedefaultmidpunct}
{\mcitedefaultendpunct}{\mcitedefaultseppunct}\relax
\EndOfBibitem
\bibitem[Herod \emph{et~al.}(2016)Herod, Chambers, and Veres]{Herod:2016}
T.~W. Herod, N.~C. Chambers and S.~P. Veres, \emph{Acta Biomaterialia}, 2016,
  \textbf{42}, 296--307\relax
\mciteBstWouldAddEndPuncttrue
\mciteSetBstMidEndSepPunct{\mcitedefaultmidpunct}
{\mcitedefaultendpunct}{\mcitedefaultseppunct}\relax
\EndOfBibitem
\bibitem[Boote \emph{et~al.}(2003)Boote, Dennis, Newton, Puri, and
  Meek]{Boote:2003}
C.~Boote, S.~Dennis, R.~H. Newton, H.~Puri and K.~M. Meek, \emph{Investigative
  Ophthalmology {\&} Visual Science}, 2003, \textbf{44}, 2941--2948\relax
\mciteBstWouldAddEndPuncttrue
\mciteSetBstMidEndSepPunct{\mcitedefaultmidpunct}
{\mcitedefaultendpunct}{\mcitedefaultseppunct}\relax
\EndOfBibitem
\bibitem[Birk \emph{et~al.}(1990)Birk, Fitch, Babiarz, Doane, and
  Linsenmayer]{Birk:1990}
D.~E. Birk, J.~M. Fitch, J.~P. Babiarz, K.~J. Doane and T.~F. Linsenmayer,
  \emph{Journal of Cell Science}, 1990, \textbf{95 ( Pt 4)}, 649--657\relax
\mciteBstWouldAddEndPuncttrue
\mciteSetBstMidEndSepPunct{\mcitedefaultmidpunct}
{\mcitedefaultendpunct}{\mcitedefaultseppunct}\relax
\EndOfBibitem
\bibitem[med()]{medium}
Tissue-dependent biochemical parameters of the medium surrounding fibrils, such
  as the salt composition, pH, and temperature, will also affect the
  dimensioned and dimensionless fibrillar constants used in our model.\relax
\mciteBstWouldAddEndPunctfalse
\mciteSetBstMidEndSepPunct{\mcitedefaultmidpunct}
{}{\mcitedefaultseppunct}\relax
\EndOfBibitem
\bibitem[Brodsky \emph{et~al.}(1980)Brodsky, Eikenberry, and
  Cassidy]{Brodsky:1980}
B.~Brodsky, E.~F. Eikenberry and K.~Cassidy, \emph{Biochimica et Biophysica
  Acta}, 1980, \textbf{621}, 162--166\relax
\mciteBstWouldAddEndPuncttrue
\mciteSetBstMidEndSepPunct{\mcitedefaultmidpunct}
{\mcitedefaultendpunct}{\mcitedefaultseppunct}\relax
\EndOfBibitem
\bibitem[Jorge-Herrero \emph{et~al.}(1999)Jorge-Herrero, Fern{\'a}ndez, Turnay,
  Olmo, Calero, Garc{\'i}a, Freile, and Castillo-Olivares]{JorgeHerrero:1999}
E.~Jorge-Herrero, P.~Fern{\'a}ndez, J.~Turnay, N.~Olmo, P.~Calero,
  R.~Garc{\'i}a, I.~Freile and J.~L. Castillo-Olivares, \emph{Biomaterials},
  1999, \textbf{20}, 539--545\relax
\mciteBstWouldAddEndPuncttrue
\mciteSetBstMidEndSepPunct{\mcitedefaultmidpunct}
{\mcitedefaultendpunct}{\mcitedefaultseppunct}\relax
\EndOfBibitem
\bibitem[Gusachenko \emph{et~al.}(2012)Gusachenko, Tran, Houssen, Allain, and
  Schanne-Klein]{Gusachenko:2012}
I.~Gusachenko, V.~Tran, Y.~G. Houssen, J.-M. Allain and M.-C. Schanne-Klein,
  \emph{Biophysical Journal}, 2012, \textbf{102}, 2220--2229\relax
\mciteBstWouldAddEndPuncttrue
\mciteSetBstMidEndSepPunct{\mcitedefaultmidpunct}
{\mcitedefaultendpunct}{\mcitedefaultseppunct}\relax
\EndOfBibitem
\bibitem[Rou{\`e}de \emph{et~al.}(2020)Rou{\`e}de, Schaub, Bellanger, Ezan, and
  Tiaho]{Rouede:2020}
D.~Rou{\`e}de, E.~Schaub, J.-J. Bellanger, F.~Ezan and F.~Tiaho, \emph{Optics
  Express}, 2020, \textbf{28}, 4845--4858\relax
\mciteBstWouldAddEndPuncttrue
\mciteSetBstMidEndSepPunct{\mcitedefaultmidpunct}
{\mcitedefaultendpunct}{\mcitedefaultseppunct}\relax
\EndOfBibitem
\bibitem[Mosser \emph{et~al.}(2006)Mosser, Anglo, Helary, Bouligand, and
  Giraud-Guille]{Mosser:2006}
G.~Mosser, A.~Anglo, C.~Helary, Y.~Bouligand and M.-M. Giraud-Guille,
  \emph{Matrix Biology}, 2006, \textbf{25}, 3--13\relax
\mciteBstWouldAddEndPuncttrue
\mciteSetBstMidEndSepPunct{\mcitedefaultmidpunct}
{\mcitedefaultendpunct}{\mcitedefaultseppunct}\relax
\EndOfBibitem
\bibitem[Ker(1981)]{Ker:1981}
R.~F. Ker, \emph{Journal of Experimental Biology}, 1981, \textbf{93},
  283--302\relax
\mciteBstWouldAddEndPuncttrue
\mciteSetBstMidEndSepPunct{\mcitedefaultmidpunct}
{\mcitedefaultendpunct}{\mcitedefaultseppunct}\relax
\EndOfBibitem
\bibitem[Straley(1973)]{Strayley:1973}
J.~Straley, \emph{Physical Review A}, 1973, \textbf{8}, 2181\relax
\mciteBstWouldAddEndPuncttrue
\mciteSetBstMidEndSepPunct{\mcitedefaultmidpunct}
{\mcitedefaultendpunct}{\mcitedefaultseppunct}\relax
\EndOfBibitem
\bibitem[Ferrarini(2010)]{Ferrarini:2010}
A.~Ferrarini, \emph{Liquid Crystals}, 2010, \textbf{37}, 811--823\relax
\mciteBstWouldAddEndPuncttrue
\mciteSetBstMidEndSepPunct{\mcitedefaultmidpunct}
{\mcitedefaultendpunct}{\mcitedefaultseppunct}\relax
\EndOfBibitem
\bibitem[Kadler \emph{et~al.}(1987)Kadler, Hojima, and Prockop]{Kadler:1987}
K.~E. Kadler, Y.~Hojima and D.~Prockop, \emph{Journal of Biological Chemistry},
  1987, \textbf{262}, 15696--15701\relax
\mciteBstWouldAddEndPuncttrue
\mciteSetBstMidEndSepPunct{\mcitedefaultmidpunct}
{\mcitedefaultendpunct}{\mcitedefaultseppunct}\relax
\EndOfBibitem
\bibitem[Jawerth \emph{et~al.}(2018)Jawerth, Ijavi, Ruer, Saha, Jahnel, Hyman,
  J{\"u}licher, and Fischer-Friedrich]{Jawerth:2018}
L.~M. Jawerth, M.~Ijavi, M.~Ruer, S.~Saha, M.~Jahnel, A.~A. Hyman,
  F.~J{\"u}licher and E.~Fischer-Friedrich, \emph{Physical Review Letters},
  2018, \textbf{121}, 258101\relax
\mciteBstWouldAddEndPuncttrue
\mciteSetBstMidEndSepPunct{\mcitedefaultmidpunct}
{\mcitedefaultendpunct}{\mcitedefaultseppunct}\relax
\EndOfBibitem
\bibitem[Misof \emph{et~al.}(1997)Misof, Rapp, and Fratzl]{Misof:1997}
K.~K. Misof, G.~G. Rapp and P.~P. Fratzl, \emph{Biophysical Journal}, 1997,
  \textbf{72}, 1376--1381\relax
\mciteBstWouldAddEndPuncttrue
\mciteSetBstMidEndSepPunct{\mcitedefaultmidpunct}
{\mcitedefaultendpunct}{\mcitedefaultseppunct}\relax
\EndOfBibitem
\end{mcitethebibliography}
\bibliographystyle{rsc} 
\end{document}